\documentclass[a4paper,12pt]{article}
\usepackage{jcappub}
\usepackage{bm}

\usepackage{graphicx}
\usepackage{newtxmath,newtxtext,amsfonts}
\usepackage[caption=false]{subfig}

\usepackage{textcomp}

\usepackage{float}

\usepackage{hyperref}
\hypersetup{
	colorlinks=true,
	linkcolor=blue,
	filecolor=violet,     
	urlcolor=blue,
	citecolor=red
}
\usepackage[]{natbib}
\usepackage{xcolor}
\usepackage{orcidlink}
\usepackage{epsfig}
\usepackage{caption}

\usepackage[toc]{appendix}
\usepackage{commath}
\usepackage{cancel}
\usepackage{csquotes}
\usepackage{placeins}
\usepackage{graphicx}

\usepackage{float}

\usepackage{multirow}
\usepackage{tcolorbox}

\newcommand{\bq}{\begin{equation}}
	\newcommand{\eq}{\end{equation}}
\newcommand{\bqn}{\begin{eqnarray}}
	\newcommand{\eqn}{\end{eqnarray}}
\newcommand{\nb}{\nonumber}
\newcommand{\lb}{\label}

\newcommand{\om}{\ensuremath{\Omega_{M}}}
\newcommand{\ct}{\ensuremath{c_{T}^{2}}}
\newcommand{\mo}{\ensuremath{\mathcal{M}_{0}}}
\newcommand{\ti}[1]{\ensuremath{\tilde{#1}}}

\title{Interacting Models of Dark Energy and Dark Matter in Einstein scalar Gauss Bonnet Gravity}

\author[1]{Saddam Hussain}
\emailAdd{mdsaddamh6@gmail.com}
\affiliation[1]{Department of Physics, Indian Institute of Technology Kanpur, Uttar Pradesh 208016, India}

\author[2]{Simran Arora}
\emailAdd{dawrasimran27@gmail.com}
\affiliation[2]{Department of Mathematics, CDOE, Chandigarh University, Punjab, 140413, India}

\author[3]{Yamuna Rana}
\emailAdd{Yamuna$\_$Rana1@baylor.edu}
\affiliation[3]{GCAP-CASPER,  Department of Physics {and Astronomy}, Baylor University,  One Bear Place $\#$97316, TX 76798-7316, USA}

\author[3]{Benjamin Rose}
\emailAdd{Ben$\_$Rose@baylor.edu}
\affiliation[3]{Department of Physics {and Astronomy}, Baylor University, One Bear Place $\#$97316, Waco, TX 76798-7316, USA}

\author[4]{Anzhong Wang}
\emailAdd{Anzhong$\_$Wang@baylor.edu; the corresponding author}
\affiliation[4]{GCAP-CASPER,  Department of Physics {and Astronomy}, Baylor University,  One Bear Place $\#$97316, TX 76798-7316, USA}

\date{\today}

\abstract{We study the dynamics of the interacting models between the Gauss-Bonnet (GB) coupled scalar field and the dark matter fluid in a homogeneous and isotropic background. A key feature of GB coupling models is the varying speed of gravitational waves (GWs). We utilize recent constraints on the GW speed and conduct our analysis in two primary scenarios: model-dependent and model-independent. In the model-dependent scenario, where determining the GW speed requires a specific GB coupling functional form, we choose an exponential GB coupling. We adopt a dynamical system analysis to obtain the necessary constraints on the model parameters that describe different phases of the universe and produce a stable late-time accelerating solution following the GW constraint, and find that to satisfy all these constraints, fine-tuning of the free parameters involved in the models is often needed. In the model-independent scenario, the GW speed is fixed to one, and we construct the autonomous system to identify the late-time stable accelerating critical points. Furthermore, we adopt a Bayesian inference method using late-time observational data sets, including 31 data points from cosmic chronometer data (Hubble data) and 1701 data points from Pantheon+ and find that all the observational constraints can be satisfied without fine-tuning. In addition, we also utilize simulated binned Roman and LSST data to study the evolution of the universe in the model-independent scenario. We find that the model shows significant deviation at higher redshifts from $\Lambda$CDM and fits the current data much better than $\Lambda$CDM within the error bars.}

\begin{document}
	\flushbottom
	\maketitle
	
	\section{Introduction}
	
	In the era of precision cosmology, our comprehension of the evolution of universe has significantly progressed. Remarkably, the High Redshift Supernova Team and the Supernova Cosmology Project discovered that distant Type Ia supernovae (SNeIa) were receding at an accelerating rate \cite{SupernovaCosmologyProject:1998vns,SupernovaSearchTeam:1998fmf}. Subsequent evidence from the Cosmic Microwave Background (CMB), baryon acoustic oscillations (BAO), and large-scale structure surveys also confirmed this accelerating expansion \cite{WMAP:2003elm,Sherwin:2011gv,Wright:2007vr,DES:2016qvw,DES:2021esc,SDSS:2005xqv}. This observation led to the concept of dark energy (DE), which is thought to drive this phenomenon. However, its true nature remains a mystery. Consequently, there is increasing interest in various cosmological models to explain these observations. 
	
	The flat $\Lambda$CDM model is generally consistent with observations, yet some recent data suggest possible discrepancies. However, the model also faces serious theoretical issues, known as the cosmological constant problem and the coincidence problem \cite{Bull:2015stt,Perivolaropoulos:2021jda,Mazumdar:2021yje,Weinberg:1988cp,Bengochea:2019daa}. The observational discrepancies include variations in the measured values of the Hubble constant $H_0$ and the amplitude of matter fluctuations $\sigma_8$ when different methods are used \cite{Kamionkowski:2022pkx,Poulin:2022sgp}. In particular,
	it has shown distinct values for the Hubble parameter: \(H_0 = 73.2\pm 1.3 \ \rm km/s/Mpc\) when measured in the low-redshift regime \cite{Riess:2020fzl}, while high-redshift measurements, such as those from CMB, yield \(H_0 = 67.4\pm0.5 \ \rm km/s/Mpc\) \cite{Planck:2018vyg}. This discrepancy, approximately at the 4.2 \(\sigma\) level, gives rise to the cosmological tension problem. Additionally, some anomalies arise when comparing theoretical predictions of the model, based on best-fit cosmological parameters, with actual observations \cite{Peebles:2022akh}. These potential discrepancies encourage the exploration of extensions to the flat $\Lambda$CDM model. 
	
	A significant number of alternatives have been proposed, where the so-called cosmological constant \(\Lambda\) is replaced with dynamical DE models. Examples include scalar field candidates such as the quintessence field \cite{Peebles:2002gy,Nishioka:1992sg,Ferreira:1997hj,Zlatev:1998tr,Copeland:1997et,Hebecker:2000au,Hebecker:2000zb,Roy:2022fif,Hussain:2023kwk,Das:2023rat,Yang:2018xah}, the k-essence field \cite{Armendariz-Picon:2000nqq,Armendariz-Picon:2000ulo,Chiba:1999ka,Fang:2014qga,Armendariz-Picon:1999hyi,Armendariz-Picon:2005oog,Arkani-Hamed:2003pdi,Scherrer:2004au,Chatterjee:2021ijw,Hussain:2022osn,Bhattacharya:2022wzu}, and the generalized Chaplygin gas model \cite{Bento:2002ps}. Other alternatives focus on modification of gravity models \cite{Clifton:2011jh,Wang:2017brl,Nojiri:2017ncd,Ishak:2018his,Shankaranarayanan:2022wbx}, for instances \(f(R)\) gravity \cite{Sotiriou:2008rp,Nojiri:2006ri,Wolschin:2010skf,Lue:2003ky,Kunz:2006ca,Wang:2022xdw,Jain:2007yk}, \(f(T)\) gravity \cite{Yang:2010hw,Cai:2015emx}, \(f(Q)\) gravity \cite{Anagnostopoulos:2021ydo,Sokoliuk:2023ccw} are few examples. Among them, the scalar-tensor theories stand out, because both their simplicity
	and consistency with observations \cite{Fujii_Maeda_2003,Ishak:2019aay}. Imposing that: (1) the theories be general covariant and (2) the field equations be second-order in terms of both the metric and the scalar field, it turns out that the Horndeski theories \cite{1974IJTP...10..363H,Horndeski:2024sjk} are the unique ones \cite{Kobayashi:2019hrl}. In this content, DE models have been extensively studied \cite{Kase:2018aps}, and also extended to models beyond Horndeski theories \cite{Langlois:2018dxi}.

	In this paper, we study  DE in a large class of  Horndeski theories, {\em Einstein-scalar-Gauss-Bonnet (EsGB) gravity} \cite{Lovelock1971TheET}, which are well-motivated by string/M-theory \cite{Gross:1986mw,Bento:1995qc,Ferrara:1996hh,Antoniadis:1997eg}. The theory is {self-consistent (such as free of ghosts and Laplacian stability)} \cite{Feir:2024,Tsujikawa:2022aar,Fernandes:2022zrq,Koivisto:2006xf,DeFelice:2009rw,DeFelice:2006pg,Tsujikawa:2006ph} \footnote{{Solar system tests restricted $\left|f'(\phi_0)\right| \lesssim 1.6 \times 10^{14}\; km^2$ \cite{Amendola:2007ni}, but the observations of low-mass x-ray binary and gravitational waves imposed much more severe  constraint $\left|f'(\phi_0)\right| \lesssim 1.18\; km^2$ \cite{Yagi:2012gp,Lyu:2022gdr}, where $\phi_0$ is the current value of the scalar field $\phi$. In addition, the constraints found in Refs. \cite{PhysRevLett.119.251304,PhysRevLett.121.221101,PhysRevLett.122.061301} do not apply to EsGB gravity \cite{Feir:2024}. In particular, for $c_T = 1$, the theory is free of ghosts and Laplacian instability in all epochs of the Universe, as long as $\dot{f}(\phi_0) \ge 0$, where an over dot denotes the derivative with respect to the cosmic time. For the detail, see  \cite{Feir:2024}.}}, and have been studied recently in various contents (See, for example, \cite{Hwang:2005hb,Wu:2017joj,Doneva:2017bvd,Li:2022xww,Pombo:2023lxg,Almeida:2024uph,Minamitsuji:2024twp}  and references therein).
	In EsGB gravity, the scalar field is coupled with the  Gauss-Bonnet (GB) term,  \(\mathcal{G} \equiv R^2 - 4R_{\mu\nu}R^{\mu\nu} + R_{\mu\nu\rho\sigma}R^{\mu\nu\rho\sigma}\), through the form
	$\alpha f(\phi)\mathcal{G}$, where $\alpha$ represents the coupling straight, and $f(\phi)$ is a function of the scalar field $\phi$ only. When $f(\phi) \propto e^{\beta \phi}$, it corresponds to the Einstein-dilaton-Gauss-Bonnet gravity \cite{Lyu:2022gdr}, and when $f(\phi) \propto \phi$, the corresponding theory has a shift  symmetry $\phi \rightarrow \phi + \phi_0$ \cite{Yagi:2011xp}. 

	The GB term couples to the scalar field \(\phi\), resulting in a non-zero contribution to the Friedmann equation, which
	has been investigated in detail in the contexts of inflation \cite{Guo:2010jr,Jiang:2013gza,Rashidi:2020wwg,Zhang:2023lrz}, {as a possible resolution to the Hubble tension \cite{Wang:2021kuw}, and  DE \cite{Carter:2005fu,Tsujikawa:2006ph,PhysRevD.71.123509,PhysRevD.75.086002,Leith:2007bu,PhysRevD.75.023518,PhysRevD.79.103008,Alimohammadi:2009js,PhysRevD.90.123512,Odintsov:2019clh,Zhang20a,Odintsov:2022zrj,Nojiri:2023mvi,Odintsov:2023aaw,Nojiri:2023mbo,TerenteDiaz:2023iqk}.} A generic feature of GB-coupled scalar field models is the non-constant speed of gravitational waves (GWs). However, recent detections of GWs emitted by the merger of two neutron stars, along with corresponding electromagnetic detections, impose stringent constraints on the GW speed. The delay between the GW and \(\gamma\)-ray detections led to the bound \(|\ct- 1| < 5 \times 10^{-16}\), where $c_T$ is the GW speed expressed in natural units \cite{LIGOScientific:2017zic,Odintsov:2019clh,Ezquiaga:2017ekz}.
	
	In Ref. \cite{TerenteDiaz:2023iqk}, the authors assumed the scalar field $\phi$ to describe DE and the coupling of $\phi$  with  the dark matter (DM) described by a pressureless fluid to be minimal, that is, the DM is conserved. Then, they investigated both cases $c_T \not= 1$ and $c_T = 1$. In the former, the  GB coupling  function $f(\phi)$ and the scalar field potential $V(\phi)$ were assumed to be of the exponential form, while in the latter only $V(\phi)$ was assumed to be of the exponential form, as the condition  $c_T = 1$ leads to $\dot{f}(\phi(t)) = C_0 a(t)$, which is sufficient to study the resulted dynamical system without specifying the exact form of $f(\phi)$, where $C_0$ is a postive constant.  In the case $c_T \not= 1$,   an accelerating, stable attractor solution following the matter-dominated phase was found. However, it was shown that the corresponding model does not satisfy the GW constraint, \(|\ct- 1| < 5 \times 10^{-16}\). In particular, in this sepcial case the authors found that the de Sitter space is no longer an attractor but a saddle fixed point. Then, the corresponding model cannot describe our current Universe either.

	In the present work, we consider the same problem as that studied in \cite{TerenteDiaz:2023iqk}, but drop the assumption of ``minimally coupling" between DE and DM, so that the energy-momentum tensor of the DM fluid is no longer conserved, \(\nabla_{\mu}T^{\mu \nu}_{\rm DM} = Q^{\nu}\), where \(Q^{\nu}\) is the interaction term that ensures the flow of energy between DM and  the scalar field $\phi$ \cite{Salvatelli:2014zta,Yang:2017ccc,Ferreira:2014jhn,Wang:2016lxa,Wang:2024vmw}. We choose the interaction to be proportional to the matter energy density and field density \(\dot{\phi}\), and explore the dynamics using dynamical system stability analysis \cite{Coley:2003mj,Rendall:2001it,Boehmer:2011tp,Dutta:2017wfd,blackmore2011nonlinear,Bouhmadi-Lopez:2016dzw,elias2006critical,Bahamonde:2017ize}. In the case $c_T \not= 1$, we find that the model exhibits an attractor accelerating solution followed by a scaling solution, and satisfies the GW constraint. However, to satisfy the GW constraint, the initial conditions corresponding to the GB coupling parameter must be fine-tuned.
	%
	%
	In the case $c_T = 1$,  we study  the corresponding dynamical system in detail  for the scalar field potential also taking an  exponential form, and   determine the constraints on the model parameters. The model yields a stable accelerating phase depending on the potential parameter and coupling constant. We also constrain the model parameters using 31 Cosmic Chronometer Hubble data and 1701 Pantheon+ samples, and find that these observational constraints can be satisfied by properly choosing the parameters involved in the solutions. 
	
	{It should be noted that in the studies of DE in EsGB gravity it has been mainly focused on models without interaction between DE and  DM \cite{Carter:2005fu,PhysRevD.71.123509,PhysRevD.75.086002,Leith:2007bu,PhysRevD.75.023518,PhysRevD.79.103008,Alimohammadi:2009js,PhysRevD.90.123512,Odintsov:2019clh,Zhang20a,Odintsov:2022zrj,Nojiri:2023mvi,Odintsov:2023aaw,Nojiri:2023mbo,TerenteDiaz:2023iqk}, except for \cite{Tsujikawa:2006ph}, in which various viable models were found. However, since the investigation was carried out before the observation of the gravitational wave, GW170717, the constraint \(|\ct- 1| < 5 \times 10^{-16}\) was not considered. As this is a severe constraint, it highly restricts the physical viable phase space of the EsGB gravity, as to be shown in this paper.}
	
	In addition, we also utilize the binned Roman and LSST data to illustrate the current model evolution (not for constraining the model parameters) compared to the standard model, $\Lambda$CDM. The current model shows significant deviations at higher redshifts and has a better fit than the fiducial cosmological model $\Lambda$CDM  within the error bars.
	
	The paper is structured as follows: In Section \ref{sec:theoretical_framework}, we discuss the theoretical framework for the coupled EsGB  gravity, and the corresponding dynamics for $c^2_T \not= 1$ are presented in Section \ref{sec:dyn_expo_int}. A constrained dynamical analysis corresponding to \(\ct = 1\) is discussed in Section \ref{sec:const_dyn}. The detailed data analysis for \(\ct = 1\) is presented in Section \ref{sec:MCMC}. Finally, a brief conclusion is outlined in Section \ref{sec:conclusion}.
	
	Before proceeding to the next section, we would like also to note that Horndeski theories reduce to EsGB gravity by properly choosing the four arbitrary functions
	$G_I(\phi, X),\; (I = 2, 3, 4, 5)$, appearing in the action, where $X \equiv - (\partial_{\mu}\phi)^2/2$. In particular,
	$G_5(\phi, X)$ must be chosen as $G_5(\phi, X) = -4\alpha f'(\phi)\ln X$  \cite{Kobayashi:2019hrl,Kase:2018aps}. However, to satisfy the GW constraint,
	\(|\ct- 1| < 5 \times 10^{-16}\), one usually sets $G_5 = 0$ (See for example,  \cite{Kubota:2022lbn} and references therein), which leads to $f(\phi) =$ Constant, and then the GB term becomes  topological, and the corresponding EsGB gravity reduces to general relativity.


	\section{Coupled Einstein-scalar-Gauss-Bonnet gravity}
	\label{sec:theoretical_framework}

	In the framework of  
	EsGB gravity, we shall consider the viability of interacting models of DE and DM. The action of EsGB gravity takes the form \cite{Zhang20a}
	\bqn
	\lb{eq2.1}
	S_{\text{EsGB}} &=& \int dx^4\sqrt{-g}\Big[ \frac{1}{2\kappa} (R-2\Lambda)  + f(\phi){\cal G} +{\cal L}_{\phi}\left(\phi\right) +  {\cal L}_{m}\left(g_{\mu\nu}, \phi; \psi\right)\Big], ~~~
	\eqn
	where $\kappa  \equiv 8\pi G/c^4$, with $G$ being the Newtonian constant, $c$ the speed of light in vacuum, and $g \;[\equiv \text{det}(g_{\mu\nu})]$. The scalar field $\phi$ is non-minimally coupled to the GB term ${\cal G}$ through the arbitrary function $f(\phi)$ with a coupling constant $\alpha$, and
	$R$ and $\Lambda$ denote the Ricci scalar and the cosmological constant,  respectively. While  ${\cal L}_{\phi}$ is that for  the scalar field with \cite{Zhang20a}
	\bqn
	\lb{eq2.2}
	{\cal L}_\phi&=&-\frac{1}{2}\left(D^\mu\phi\right)\left(D_\mu\phi\right)-V(\phi),\nonumber\\
	{\cal G}&=&R^2 -4R_{\mu\nu}R^{\mu\nu} +R_{\mu\nu\rho\sigma}R^{\mu\nu\rho\sigma},
	\eqn
	where $V(\phi)$ is the potential of the scalar field, and $D_{\mu}$ denotes the covariant derivative with respect to the metric $g_{\mu\nu}$. The  Lagrangian density  ${\cal L}_{m}$ represents
	the matter fields  filled in our Universe, including DM and collectively denoted by $\psi$. In this paper, we shall consider the scalar field $\phi$ to describe DE. 
	From the above action, one can derive the equations of motion for both $g_{\mu\nu}$ and $\phi$,
	\bqn
	\label{eq2.3a}
	&& R_{\mu\nu}-\frac{1}{2}g_{\mu\nu}R + \Lambda g_{\mu\nu}= T^{GB}_{\mu\nu}+ \kappa T^\phi_{\mu\nu} +\kappa T^m_{\mu\nu}, ~~~\\
	\label{eq2.3b}
	&& D^2\phi= \frac{dV}{d\phi}-\alpha \frac{d f}{d\phi}{\cal G} - Q,
	\eqn
	where $D^2 \equiv g^{\mu\nu}D_{\mu}D_{\nu}$, and 
	\bqn
	\lb{eq2.4}
	T^{GB}_{\mu\nu}&\equiv&2\left(D_\mu D_\nu f\right)R-2g_{\mu\nu}\left(D_\rho D^\rho f\right)R 
	-4\left(D^\rho D_\nu f\right)R_{\mu\rho}-4\left(D^\rho D_\mu f\right)R_{\nu\rho}\nonumber\\
	&&+4\left(D^\rho D_\rho f\right)R_{\mu\nu}+4g_{\mu\nu}\left(D^\rho D^\sigma f\right)R_{\rho\sigma}
	-4\left(D^\rho D^\sigma f\right)R_{\mu\rho\nu\sigma}\ ,\nonumber\\
	T^\phi_{\mu\nu}&\equiv&\frac{1}{2}\left(D_\mu\phi\right) D_\nu\phi-\frac{1}{4} g_{\mu\nu}\Big[\left(D^\alpha\phi\right)D_\alpha\phi + 2V(\phi)\Big],\nb\\
	T^m_{\mu\nu} &\equiv& - \frac{2}{\sqrt{-g}}\frac{\delta\left(\sqrt{-g}\; {\cal{L}}_{m}\right)}{\delta g^{\mu\nu}},\nb\\
	Q &\equiv& 2\kappa \frac{\delta{\cal{L}}_{m}}{\delta\phi}.
	\eqn
	
	In the flat FLRW background \(ds^2 = -dt^2 + a(t)^2 d\vec{x}^2\), where \(a\) is the scale factor, the Hubble equation is expressed as
	\begin{equation}
		3 H^2  =  \rho_{m} + \rho_{\phi} + 24 f'(\phi) \dot{\phi} H^3.
		\label{frd_eqn}
	\end{equation}
	Throughout this study, we will omit considerations of the cosmological constant \(\Lambda\). According to the scalar field Lagrangian in Eq.~\eqref{eq2.2}, the energy density of the scalar field is given by \(\rho_{\phi} = \frac{1}{2}\dot{\phi}^2 + V(\phi)\), while the energy density of the fluid is denoted by \(\rho_{m}\). The equations of motion pertaining to the matter components are as follows:
	\begin{eqnarray}
		\dot{\rho}_{m} + 3 H \rho_{m} &= & Q \rho_{m} \dot{\phi}, \label{dm_eqs}\\
		\ddot{\phi} + V_{,\phi} + 3 H \dot{\phi} &=& - Q \rho_{m} - 24 f' H^2 (H^2 + \dot{H}).\label{field_eos}
	\end{eqnarray}
	Here, \(Q\) is the coupling constant and  \(\dot{()} \equiv d()/dt\) and \(()' \equiv d()/d\phi\). 
	
	\section{Dynamical Stability Analysis}
	\label{sec:dyn_expo_int}
	
	This section aims to evaluate the dynamical characteristics of the quintessence field, particularly focusing on the exponential potential with an exponential type of Gauss-Bonnet coupling parameter:
	\begin{equation}
		V(\phi) =V_0 e^{- \lambda \phi}, \quad f'(\phi) = \alpha e^{\mu \phi},
	\end{equation}
	where $V_0,\; \lambda, \alpha$ and $\mu$ are free parameters. To scrutinize the dynamics inherent in the non-minimal coupling system, we establish a set of dimensionless dynamical variables as follows: 
	\begin{equation}
		x = \dfrac{\dot{\phi}}{\sqrt{6} H}, \ y= \dfrac{\sqrt{V}}{\sqrt{3} H}, \ z = f' H^2, \ \om = \dfrac{\rho_{m}}{3 H^2}, \ \Omega_{\phi} = \dfrac{\rho_{\phi}}{3 H^2}.
	\end{equation}
	Corresponding to these variables, the Hubble equation imposes constraints on the dark matter energy density parameter as:
	\begin{equation}
		\om = 1 - \Omega_{\phi} - \Omega_{\rm GB}. \label{constrained_equation}
	\end{equation}
	As the energy components of the physical entity must remain positive, the system adheres to the following constraints:
	\begin{equation}
		0 \le \om \le 1, \quad \Omega_{\phi} >0, \quad 0<\Omega_{\phi} + \Omega_{\rm GB}<1.
		\label{constraint}
	\end{equation} 
	Here, \(\Omega_{\phi} \equiv x^2 + y^2\) and $\Omega_{\rm GB} \equiv 8 \sqrt{6} z x$. We can further express $\dot{H}/H^2$ as,
	\begin{equation}
		\frac{\dot{H}}{H^2} = \frac{12 Q \om z+24 \mu  x^2 z-3 x^2-12 \sqrt{6} x z-4 \sqrt{6} x z+12 \lambda  y^2 z-\frac{3 \om}{2}-96 z^2}{-8 \sqrt{6} x z+96 z^2+1}.\label{dotH}
	\end{equation}
	Then, the autonomous system of equations can be formulated as follows:
	\begin{eqnarray}
		x' =\frac{\dot{x}}{H} & = & -\frac{\dot{H}}{H^2} \left(x+\frac{24 z}{\sqrt{6}}\right)-\frac{3 Q \om}{\sqrt{6}}-3 x+\frac{3 \lambda  y^2}{\sqrt{6}}-\frac{24 z}{\sqrt{6}}, \label{x_prime}\\
		y' = 	\frac{\dot{y}}{H} & = & \frac{-1}{2} \sqrt{6} \lambda  x y - y \frac{\dot{H}}{H^2}, \label{y_prime}\\
		z' =	\frac{\dot{z}}{H} & = &  \sqrt{6} \mu  x z+ 2 z \frac{\dot{H}}{H^2}, \label{z_prime}
	\end{eqnarray}
	but now \(()'={d()}/({H dt})\) and \(dN =  H dt = d\ln a\). The dimension of the autonomous system is 3D; thus, the corresponding phase space is also 3D. An additional property of this phase space is the presence of an invariant submanifold in the autonomous equations for \(y'\) and \(z'\). Under the transformations \(y \to -y\) and \(z \to -z\), the equations in \(y'\) and \(z'\) remain invariant. 
	
	The dynamics of this complex system are elucidated by identifying the critical points of the autonomous equations. By evaluating these critical points, we ascertain the system's stability. Based on physical viability, constraints are imposed on the system. It's pertinent to note that critical points yielding non-physical solutions or violating the mentioned constraints will be disregarded. Since the autonomous equations exhibit an invariant submanifold in \((y,z)\), critical points with \(y \ge 0\) and \(z \ge 0\) or \(z \le 0\) will be considered. To distinguish between accelerating and non-accelerating features of the points, the effective equation of state of the composite system can be expressed as:
	\begin{equation}
		\omega_{\rm eff} = -1 - \frac{2}{3} \dfrac{\dot{H}}{H^2}.
	\end{equation}
	For the accelerating solution, the range \(-1 \leq \omega_{\rm eff} \leq -1/3\) indicates quintessential behavior, while \(\omega_{\rm eff} < -1\) exhibits phantom nature. In contrast, for the non-accelerating solution, the effective equation of state lies in \(\omega_{\rm eff} \ge 0\). The critical points corresponding to this system are summarized in Tab. \ref{tab:critical_points_quintessence}. Here, the density parameters and effective equation of state are explicitly evaluated. Additionally, the stability of the critical points is mentioned in the table by evaluating the eigenvalues. To obtain the eigenvalues corresponding to the autonomous equations, we linearize the right-hand sides up to the first order at the critical points and construct the Jacobian matrix \(J_{ij} \equiv \left.\left({d\bm{x}_{i}'}/{d\bm{x}_{j}}\right)\right|_{\bm{x}_{*}}\), where \(\bm{x}_{i=1,2,3} = \{x, y,z\}\). The eigenvalues of the matrix indicate the stability of critical points, provided that the real part of the eigenvalues is non-zero. If the real parts of critical points are negative (positive), the point becomes asymptotically stable (unstable). If the real parts contain alternating signs, the point becomes a saddle point. 
	\begin{table}[t]
		\tiny
		\centering
		\begin{tabular}{p{1.0cm}p{2.9cm}p{0.8cm}p{1.0cm}p{1.6cm} p{4.2cm}}
			\hline
			Points & \((x_*, y_*, z_*)\) & \(\Omega_{\phi}\)& $\Omega_{\rm GB}$ & $\omega_{\rm eff}$ & Eigenvalues\\
			\hline
			\hline
			$P_{1,2} $ & \((\mp 1,0,0)\) & 1 & 0& 1 & $\left(3\mp\sqrt{6} Q,3 \pm \sqrt{\frac{3}{2}} \lambda ,\sqrt{6} (\mp\mu )-6\right)$\\
			\hline
			$P_{3}$ & \(\left( -\sqrt{\dfrac{2}{3}} Q, 0, 0\right)\) & $\dfrac{2 Q^2}{3}$ & 0& $\dfrac{2 Q^2}{3}$ & $\left(Q^2-\frac{3}{2},Q (\lambda +Q)+\frac{3}{2},-2 Q (\mu +Q)-3\right)$\\
			\hline
			$P_{4}$ & $\left(\frac{\lambda }{\sqrt{6}}, \bigg| \frac{\sqrt{-\left(\left(\lambda ^2-6\right) (\lambda +Q)\right)}}{\sqrt{6} \sqrt{\lambda +Q}} \bigg|, 0\right)$ & $1$ & 0&$\frac{1}{3} \left(\lambda ^2-3\right)$ & $\left(\frac{1}{2} \left(\lambda ^2-6\right),\lambda  (\lambda +Q)-3,\lambda  (\mu -\lambda )\right)$\\
			\hline
			$P_{5}$ & \(\left(\frac{\sqrt{\frac{3}{2}}}{\lambda +Q}, \bigg| \frac{\sqrt{Q^2+\lambda  Q+\frac{3}{2}}}{\lambda +Q}\bigg| , 0\right)\) & $\frac{Q^2+\lambda  Q+3}{(\lambda +Q)^2}$ & 0& $-\frac{Q}{\lambda +Q}$ & ($\rm E_1, E_2, E_3$) -- Fig. \ref{fig:stab_p5}\\
			\hline
			$P_{6}$ & $\left(0, \frac{\sqrt{2 Q^2+1}}{\sqrt{2} \sqrt{Q} \sqrt{\lambda +Q}}, \frac{1}{16 Q}\right)$ & $\frac{2 Q^2+1}{2 Q^2+2 \lambda  Q}$ & 0 & $-1$ & ($\rm E_1, E_2, E_3$) -- Fig. \ref{fig:stab_p6} \\
			\hline
			$P_{7}$ & \((0, 1, \lambda/8)\) & 1 & 0 & -1 & \(6 \lambda  Q-3\), $-\frac{3}{2} \mp \frac{\sqrt{3} \sqrt{\left(3 \lambda ^2+2\right) \left(17 \lambda ^2-8 \lambda  \mu +6\right)}}{6 \lambda ^2+4}$\\
			\hline
			$P_{8,9}$ & -- & --&-- & -- & -- \\
			\hline
		\end{tabular}
		\caption{The critical points and their corresponding eigenvalues. Dashed lines indicate labels for reference in the text. }
		\label{tab:critical_points_quintessence}
	\end{table}
	The system produces nine critical points, among which the critical points \(P_{1}\) to \(P_{5}\) are characterized by vanishing \(z\), implying \(f' \sim 0\) for \(H \ne 0\). Despite the vanishing of the Gauss-Bonnet parameter, the coupling between the field and the fluid persists. Consequently, there exists a non-zero exchange of energy flow between dark matter and the scalar field via the coupling term \(Q\). Now, we provide a concise discussion on the nature of critical points that delineate various different phases of the system's evolution.
	
	\begin{itemize}
		
		\item \textbf{Point $P_{1,2}$:} At these points, the kinetic component of the field dominates over the potential component, and the derivative of the Gauss-Bonnet coupling factor \(f'\) approaches zero, resulting in a vanishing \(\Omega_{\rm GB}\) with a dominated scalar field density. These points feature an effective equation of state \(\omega_{\rm eff} \sim 1\), indicating an ultra-slow expansion phase, a common feature with quintessence-type scalar fields. The eigenvalues at these points depend on the model parameters \((Q, \mu, \lambda)\), and for certain choices of these parameters, the points may behave as a saddle or be unstable. Although these points do not describe the dynamics of the current phase of the universe, they may have been significant during the very early phases of the universe.
		
		\item \textbf{Point $P_{3}$:} The coordinates of this point depend on the coupling parameter \(Q\), with the kinetic component of the field dominating over its potential. The field's energy density parameter and effective equation of state are both \(Q\)-dependent, showcasing non-accelerating behavior. The real part of the eigenvalues becomes negative, indicating stability, when \(\lambda > \sqrt{6}\), \(-\sqrt{\frac{3}{2}} < Q < \frac{\sqrt{\lambda ^2 - 6}}{2} - \frac{\lambda}{2}\), and \(\mu < \frac{-2 Q^2 - 3}{2 Q}\). Beyond this region, the point becomes unstable. Similar to the previous point, it may have been significant during the very early phases when the interaction between the field and the fluid is non-zero.
		
		\item \textbf{Point $P_{4}$:} The coordinates of this point depend on the potential parameter $\lambda$ and the coupling parameter $Q$. The field's fractional density parameter dominates and becomes independent of any model parameters. From the constrained equation, Eq.~\eqref{constrained_equation}, the fluid density parameter vanishes. The effective equation of state parameter becomes \(\lambda\)-dependent and exhibits accelerating and non-accelerating solutions for \(0 < \lambda^2 < 2\) and \(3 < \lambda^2 < 6\), respectively. Corresponding to this accelerating or non-accelerating range, the point stabilizes for \(Q < \frac{3 - \lambda ^2}{\lambda}\) and \(\mu < \lambda\). Beyond this range, the point becomes unstable.
		
		\begin{figure}[t]
			\centering
			\subfloat[\label{fig:const_p5}]{\includegraphics[scale=0.6]{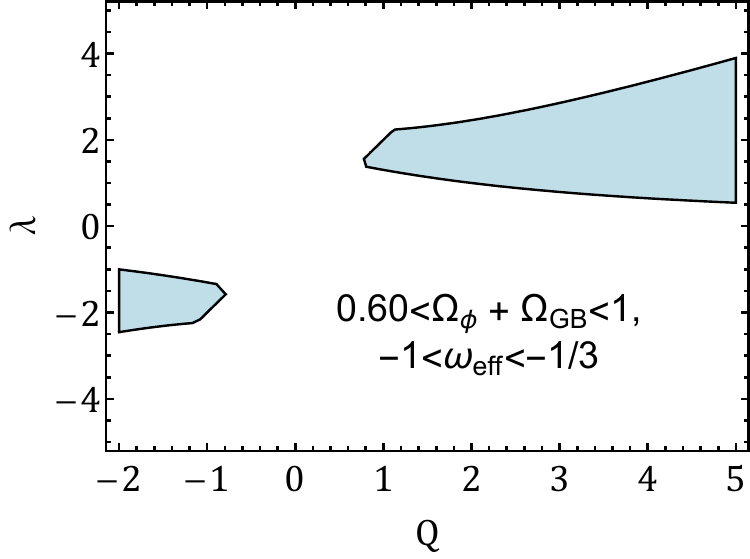} }
			\hspace{0.5cm}
			\subfloat[\label{fig:stab_p5}]{\includegraphics[scale=0.6]{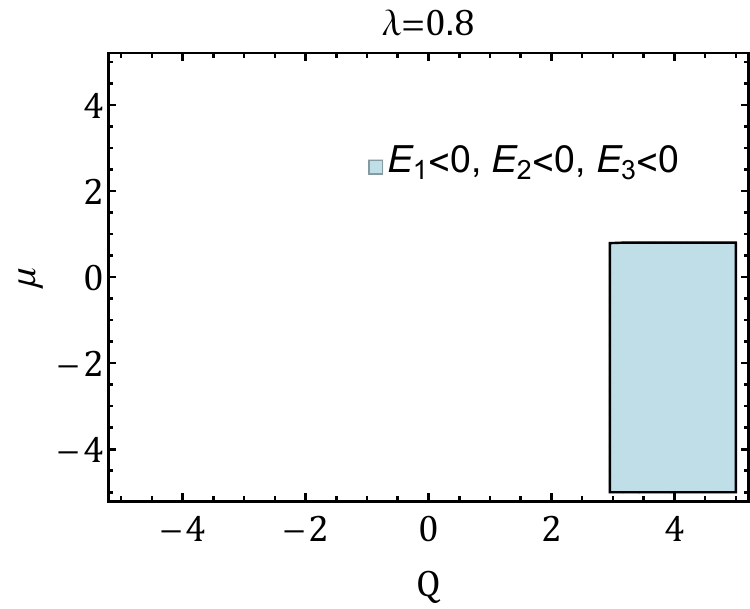}}
			\caption{(a) The energy density constrained and effective equation of state at the point $P_{5}$ in the parameter space of \((Q, \lambda)\). (b) The stability of the point for \(\lambda= 0.8\). }
		\end{figure}
		
		\item \textbf{Point $P_{5}$:} At this point, the GB coupling parameter \(z\) vanishes. The field's energy density parameter and effective equation of state depends on \(Q\) and \(\lambda\). The behavior is depicted in Fig. \ref{fig:const_p5}. In this case, we have chosen the region of the parameter space corresponding to the existence of field energy density such that if the system enters in the late-time phase, the scalar field dominates over the fluid energy density, thereby choosing the range of field's density parameter to be \(0.6 < \Omega_{\phi} + \Omega_{\rm GB} \leq 1\). 
		On evaluating stability, the eigenvalues depend on all the model parameters and take a complicated form. To discern the stability of the point, we fix one of the parameters, say the potential parameter \(\lambda = 0.8\), using the prior information obtained in Fig. \ref{fig:const_p5}. By fixing \(\lambda\), we constrain other model parameters \(\mu\) and \(Q\) for which the point exhibits stable behavior, as depicted in Fig. \ref{fig:stab_p5}.

		\item \textbf{Point $P_{6}$:} At this point, the kinetic component of the field vanishes, while the other coordinates \(y\) and \(z\) depend on \(Q\) and \(\lambda\). Although \(z\) is non-vanishing, the corresponding \(\Omega_{\rm GB}\) still vanishes. The effective equation of state becomes \(-1\), and the field's energy density becomes dependent on the model parameters \(Q\) and \(\lambda\). Similar to the previous point, we depict the behavior of \(\Omega_{\phi}\) in the parameter space of \((Q, \lambda)\) in Fig. \ref{fig:const_p6}. The eigenvalues depend on all the model parameters; hence, by fixing one of the parameters, say \(\lambda\), we obtain the stable region that constrains the rest of the model parameters \(\mu\) and \(Q\), as shown in Fig. \ref{fig:stab_p6}.
		
		\begin{figure}[t]\centering
			\subfloat[\label{fig:const_p6}]{\includegraphics[scale=0.50]{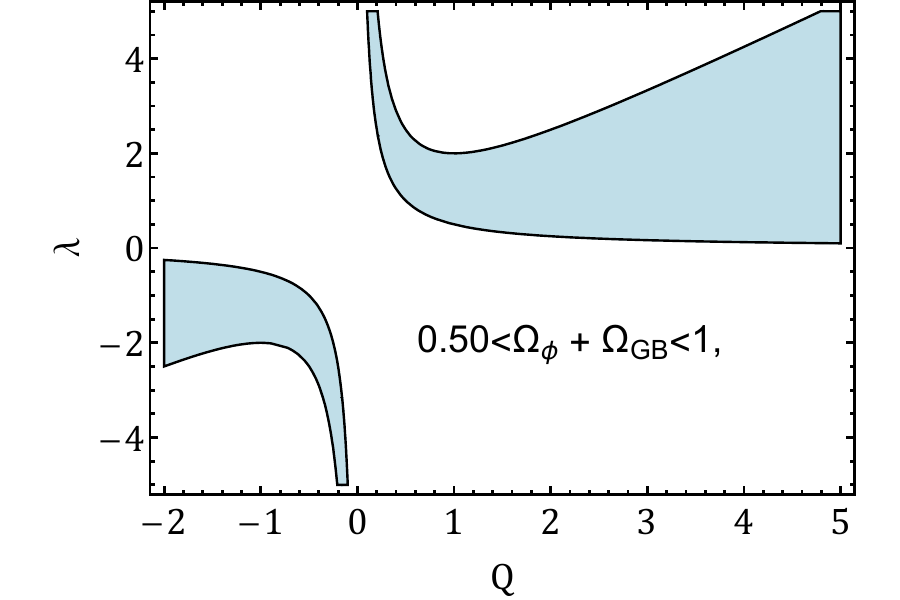} }
			\hspace{0.5cm}
			\subfloat[\label{fig:stab_p6}]{\includegraphics[scale=0.53]{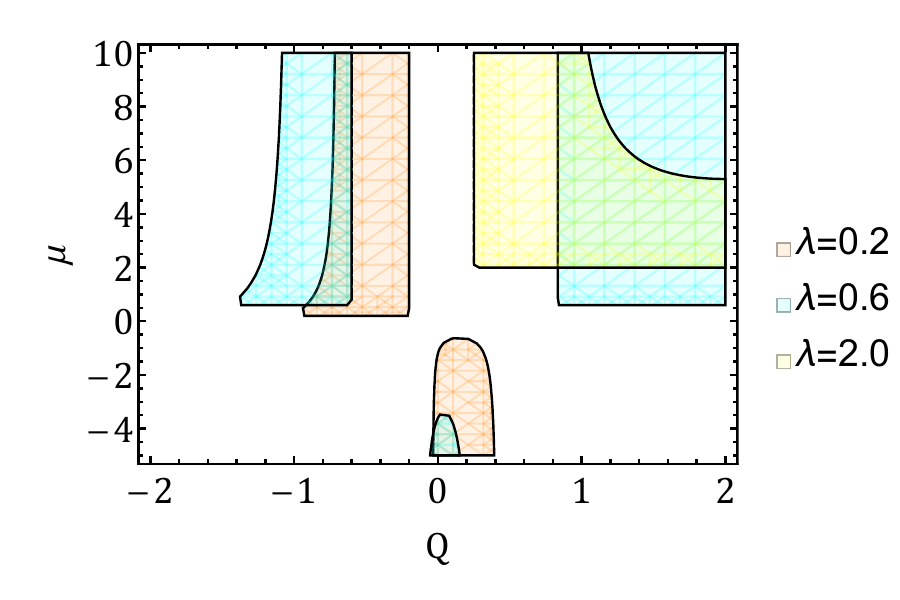}}
			\caption{(a) The energy density constraint at the point $P_{6}$ in the parameter space of \((Q, \lambda)\). (b) The stability of the point for different values of \(\lambda\). }
			
		\end{figure}
		
		\item \textbf{Point $P_{7}$:} At this point, the coordinate \(x\) vanishes, \(y\) takes a constant value, and \(z\) becomes dependent on the potential parameter \(\lambda\). The field's energy density parameter dominates over the rest of the components, and the effective equation of state becomes \(-1\). Since the coordinates of the point are independent of \(Q\), this implies that the point remains significant even though the coupling between the scalar field and dark matter fluid vanishes. The eigenvalues at this point depend on all the model parameters. Hence, by fixing \(\lambda\), we plot the region of stability in Fig. \ref{fig:stability_p7}.
		
		\begin{figure}[t]
			\centering
			\includegraphics[scale=0.5]{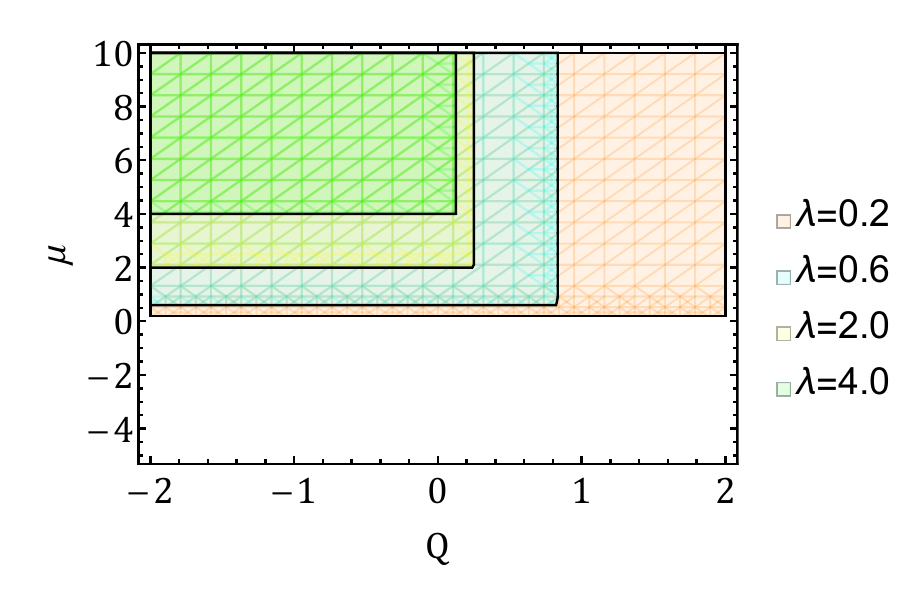}
			
			\caption{The stability of $P_{7}$ for different values of $\lambda$ in the parameter space of \((Q, \mu)\). }
			\label{fig:stability_p7}
		\end{figure}
		
		\item \textbf{Point $P_{8}$:} The coordinates of the point are given by
		\bqn
		(x_*, y_*, z_*)& =&  \left(\frac{5 \mu +\sqrt{\mu ^2-48 \mu  Q^3+\left(16-48 \mu ^2\right) Q^2-64 \mu  Q}-4 Q}{2 \sqrt{6} \mu  (\mu +Q)}, 0,\right.\nb \\
		&&~~~~~~ \left. \frac{\mu -\sqrt{\mu ^2-48 \mu  Q^3+\left(16-48 \mu ^2\right) Q^2-64 \mu  Q}+4 Q}{96 \mu  Q}\right).
		\eqn
		At this point, the fractional energy density of the field is given by 
		\bq
		\Omega_{\phi} = \frac{\left(5 \mu +\sqrt{\mu ^2-48 \mu  Q^3+\left(16-48 \mu ^2\right) Q^2-64 \mu  Q}-4 Q\right)^2}{24 \mu ^2 (\mu +Q)^2},
		\eq
		and the effective equation of state \(\omega_{\rm eff}\) and \(\Omega_{\rm GB}\) are only dependent on \((Q, \mu)\). To constrain these model parameters, we use the constraint relation given in Eq.~\eqref{constraint}, with the effective equation of state in Fig. \ref{fig:point8_nature}. In this parametric space plot, we have plotted the fluid energy density in the range \(0 < \Omega_m < 0.4\), the field sector density is plotted between \(0 <\Omega_{\phi} + \Omega_{\rm GB}<1\), and the effective equation of state in the range of \(-1<\omega_{\rm eff}<-1/3\) and \(0<\omega_{\rm eff}<1/3\). The overlapping region in blue and yellow indicates that there exists a region where the energy densities of the field sector \(( \Omega_{\phi}+\Omega_{\rm GB})\) and the fluid sector coexist. However, the green region shows the accelerating solution, which does not overlap with any other region. The red region shows the non-accelerating solution, where the point produces matter or radiation-type fluid characteristics. This region shares the overlapping region with blue and yellow, signifying that the point can exhibit only a non-accelerating solution. In the end, we compute the eigenvalues corresponding to the Jacobian matrix evaluated at this point, where one of the eigenvalues, say \(E_{3} \in \mathbb{R} < 0\), is plotted in violet color, since it is the only eigenvalue that is independent of \(\lambda\). We see that there exists no such region where all the colors overlap, implying that the point cannot exhibit a stable solution, whether accelerating or non-accelerating, as long as it follows the constraint relation Eq.~\eqref{constraint}.
		
		\begin{figure}[t]
			
			\centering
			\includegraphics[scale=1.0]{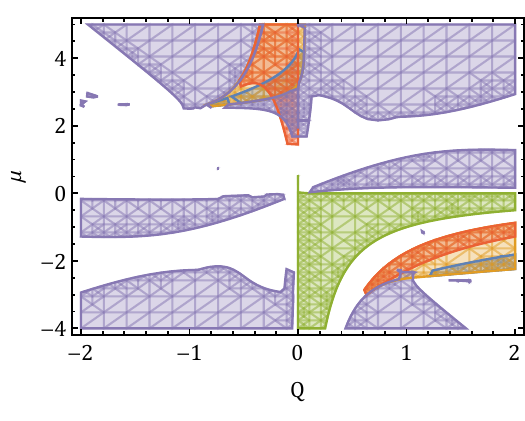}
			\caption{ The nature corresponding to point $P_{8}$, where the region colors identify the constraints on the critical point. Blue, orange, green, red, and violet colors show the constraints \(0 \le \om \le 0.4\), \(0 < \Omega_{\phi} + \Omega_{\rm GB}<1\), \(-1 < \omega_{\rm eff} < -1/3\), \(0 < \omega_{\rm eff} < 1/3\), and \(E_3 < 0\), respectively.}
			\label{fig:point8_nature}
		\end{figure}
		
		\item  \textbf{Point $P_{9}$:} The coordinates of the point are
		\bqn
		(x_*, y_*, z_*) &=& \left(\frac{5 \mu -\sqrt{\mu ^2-48 \mu  Q^3+\left(16-48 \mu ^2\right) Q^2-64 \mu  Q}-4 Q}{2 \sqrt{6} \mu  (\mu +Q)}, 0,\right.\nb\\
		&& ~~~~~~~~~~~~\left. \frac{\mu +\sqrt{\mu ^2-48 \mu  Q^3+\left(16-48 \mu ^2\right) Q^2-64 \mu  Q}+4 Q}{96 \mu  Q}\right).
		\eqn
		At this point, the field's energy density parameter is given by 
		\bq
		\Omega_{\phi} = \frac{\left(-5 \mu +\sqrt{\mu ^2-48 \mu  Q^3+\left(16-48 \mu ^2\right) Q^2-64 \mu  Q}+4 Q\right)^2}{24 \mu ^2 (\mu +Q)^2}.
		\eq
		The effective equation of state \(\omega_{\rm eff}\) and \(\Omega_{\rm GB}\) are dependent on model parameters \((Q, \mu)\).  Similar to point $P_{8}$, this point is incapable of producing an accelerating solution satisfying the constraint relation Eq.~\eqref{constraint}, as shown in Fig. \ref{fig:point9_nature}. In addition, one of the eigenvalues is dependent on \((Q, \mu)\) and the real part can yield negative values. However, there exists no overlapping region where it simultaneously satisfies the energy relation constraint and produces a stable accelerating or non-accelerating result. Thus, this point cannot be considered a viable physical point.
		\begin{figure}[t]
			\centering
			\includegraphics[scale=1.]{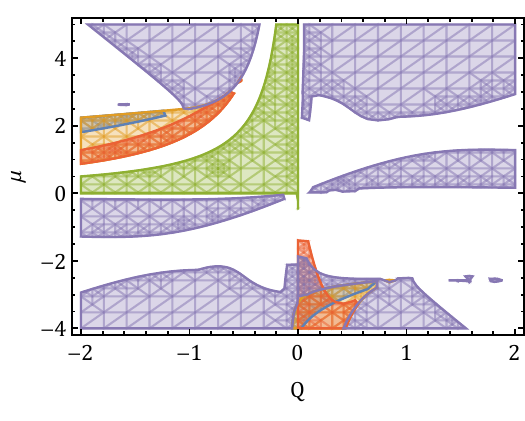}
			\caption{ The nature corresponding to point $P_{9}$, where the region colors identify the constraints on the critical point. Blue, orange, green, red, and violet colors show the constraints \(0 \le \om \le 0.4\), \(0 < \Omega_{\phi} + \Omega_{\rm GB}<1\), \(-1 < \omega_{\rm eff} < -1/3\), \(0 < \omega_{\rm eff} < 1/3\), and \(E_3 < 0\), respectively.}
			\label{fig:point9_nature}
		\end{figure}
		
	\end{itemize}
	
	After carefully studying the region of existence with the necessary constraints, it can be seen that points \(P_{6}\) and \(P_{7}\) are the viable critical points that generate an accelerating solution with dominating scalar field density, consequently describing the late-time phase of the universe. Point \(P_{6}\) depends on the interaction parameter \(Q\); hence, as long as energy flow occurs between the scalar field and fluid, this point remains significant over point \(P_{7}\).
	
	Based on the imposed constraints corresponding to point \(P_{6}\), we select the model parameters \((\mu =6, Q= 1, \lambda = 0.6)\), so that the point yields a stable and accelerating solution in the late-time phase.
	
\begin{figure}[t]
		\centering
		\subfloat[\label{fig:evo_with_coupling}]{\includegraphics[scale=0.5]{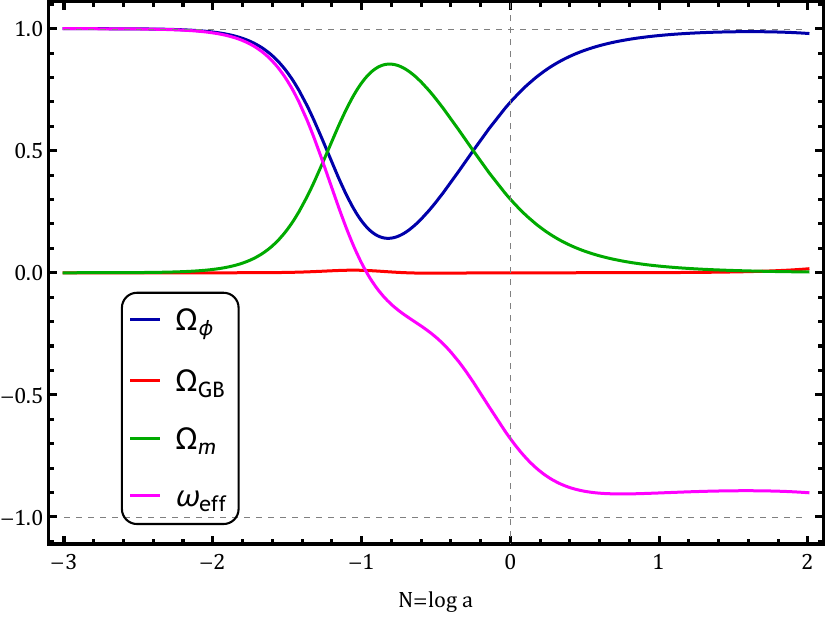}}
		\hspace{0.2cm}
		\subfloat[\label{fig:evo_with_nocoupling}]{\includegraphics[scale=0.5]{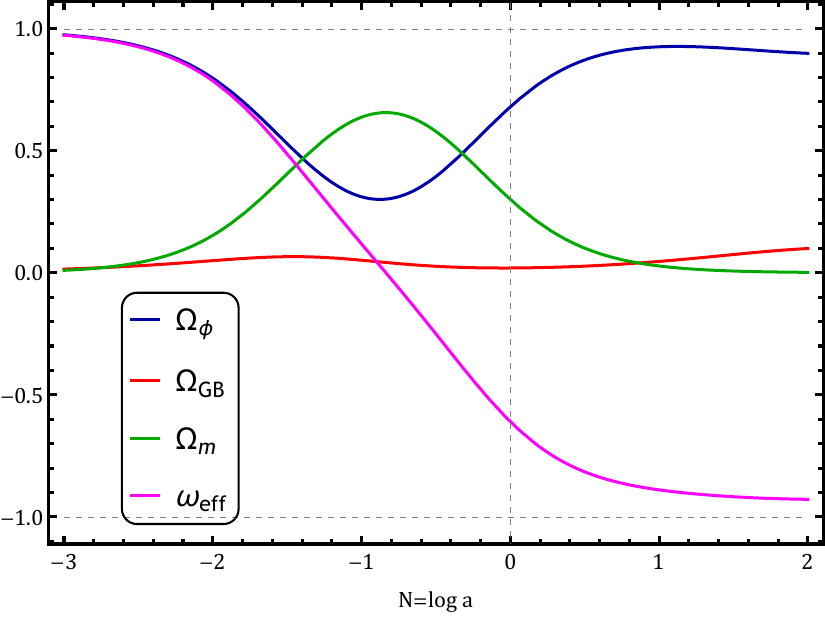}}
		\caption{ The evolution of cosmological parameter for (a) with dark matter scalar field coupling \(\mu = 6.0,\lambda = 0.6,Q= 1 \) with initial conditions \(x(0)= -0.1, y(0)= 0.83, z(0) =10^{-4}\). (b) Without coupling with \(\mu =3, \lambda=0.6, Q=0\), for the initial conditions \(x(0) = 0.2, y(0) = 0.8, z(0)= 0.005\).}
	\end{figure}

\begin{figure}[t]
		\centering
		\subfloat[\label{fig:gw_const}]{\includegraphics[scale=0.4]{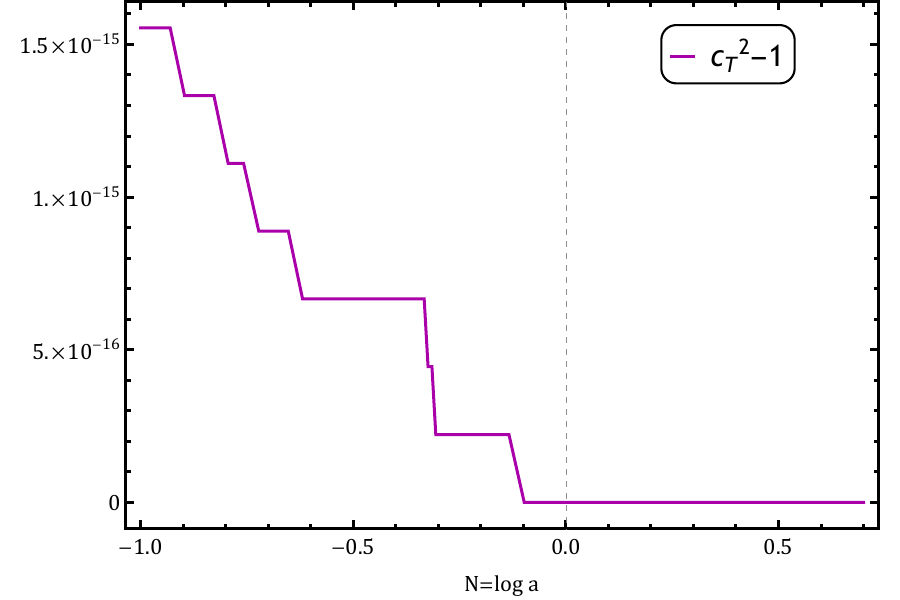}}\hspace{0.2cm}
		\subfloat[\label{fig:evo_noncoup_gw}]{\includegraphics[scale=0.4]{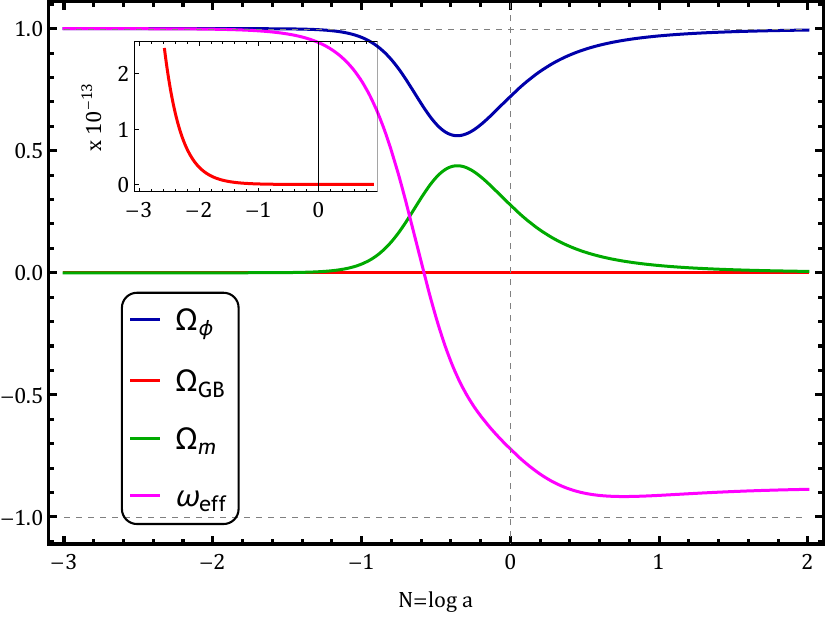}}
		\caption{ (a) The GW constraint and (b) corresponding evolution of cosmological parameters have been plotted with an initial condition \(x(0) = 0.01, y(0)=0.85, z(0)=10^{-17}\) for \(\lambda = 0.6, \mu = 1, Q = 2\). }
	\end{figure}

\begin{figure}[t]
		\centering
		\includegraphics{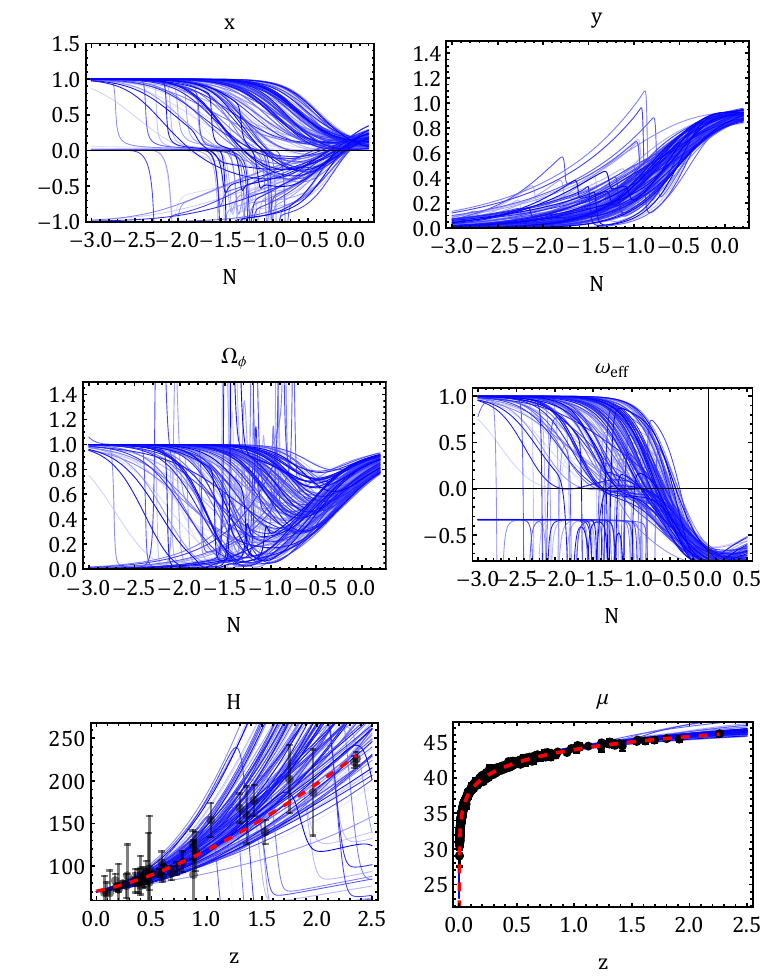}
		\caption{ The evolution of dynamical variables corresponding to the interacting case for \(H_0 = 70 \ \rm Km s^{-1} Mpc^{-1}\). Here the dotted red line is plotted for flat $\Lambda$CDM model. For $H$ vs $z$ plot, the best value corresponding to flat $\Lambda$CDM model has been used \(H_0 = 70.6  \ \rm Km \ s^{-1} Mpc^{-1},\) \(\Omega_\Lambda = 0.74\), however, for distance modulus($\mu$), \(H_0 = 73.6  \ \rm Km \ s^{-1} Mpc^{-1}\), and \(\Omega_\Lambda=0.67\).  }
		\label{fig:coupling_evo}
	\end{figure}

	The numerical evolution of the system has been shown in Fig. \ref{fig:evo_with_coupling}. It is found that for this set of benchmark points and appropriate initial conditions, the model produces a viable matter domination epoch between \(N = -2\) to \(0\), corresponding to redshift, \(z = 7\) to \(0\). Note that by selecting the benchmark points, some of the critical points may no longer exist. In Fig. \ref{fig:evo_with_nocoupling}, we have evolved the system for \(Q = 0\), indicating that the model stabilizes at \(P_{7}\).
	In addition to the constraints obtained above on the model parameters based on stability, one can further constrain the model parameters in light of the gravitational wave speed, which approximately equals the speed of light \cite{LIGOScientific:2017zic}. When considering tensor perturbations about the FLRW background metric, the tensor propagation speed \cite{Tsujikawa:2006ph,Odintsov:2019clh} becomes: 
	\begin{equation}
		\ct = \frac{1- 8 \ddot{f}}{1 - 8 H  \dot{f}}.
		\label{gw_speed}
	\end{equation}
	The propagation speed at the perturbation level must be positive to ensure the non-existence of ghost modes. However, in the context of GW170817, the ratio of tensor propagation speed to the speed of light imposes a bound: 
	\begin{equation}
		|\ct-1 | \le 5 \times 10^{-16}, \label{gw_bound}
	\end{equation}
	in the natural units where \(c= \kappa =1\) \cite{Ezquiaga:2017ekz}. Using the GB coupling factor \(f' = \alpha e^{\mu \phi}\), one can express Eq.~\eqref{gw_speed} in terms of dynamical variables as: 
	\begin{equation}
		\ct = \frac{1 - 8 \left(6 \mu z x^2 + \sqrt{6} z (x' + x \dot{H}/H^2)\right)}{1 -8 \sqrt{6}  z x }.
		\label{ct}
	\end{equation}
	On evaluating the tensor propagation speed at points \(P_{0}\) to \(P_{7}\), \(\ct\) yields \(1\). It implies that as the critical points indicate different epochs of the universe, the gravitational wave propagation speed remains equal to the speed of light at those epochs. From the stability analysis, it becomes clear that the other critical points \(P_{8}\) and \(P_{9}\) cannot produce stable accelerating solutions. However, for some choice of the parameters, these points can exist and describe the non-accelerating regime of the universe. Nevertheless, these points do not adhere to the \(\ct\) bound for any choice of \((Q, \mu)\). Hence, these points cannot be considered viable physical points.
	As we can see, after evaluating \(\ct\) at other points, it takes a constant value of \(1\). However, this may not be the case in intermediate periods. Therefore, one must choose initial conditions corresponding to the dynamical variables such that the evolution of \(\ct\) from the past to the present epoch satisfies the GW bound given in Eq.~\eqref{gw_bound}. In Fig. \ref{fig:gw_const}, we have plotted the evolution of \(\ct\) up to the scale factor for which it satisfies the above bound, and the corresponding parameter evolution has been plotted in Fig. \ref{fig:evo_noncoup_gw}. We see that to satisfy the GW bound, the variable \(z\) associated with the GB coupling parameter's initial condition must be less than \(z(0) < 10^{-15}\). We have chosen the rest of the parameter values such that the system stabilizes at \(P_{6}\) and produces an accelerating solution. However, one must vary over the possible initial conditions within the respective bound to check the evolution of the model in different epochs. Hence, to demonstrate this, we choose randomly generated initial conditions \(0.01 < x_0 < 0.2\), \(0.8 < y_0 < 0.93\), \(z_0 \sim 10^{-17}\), \(0.001 < \lambda < 1.5\), \(0 < \mu < 50\), \(0 < Q < 1.3\) within these ranges to observe the evolution of dynamical variables, the Hubble parameter, and the distance modulus \(\mu\), as shown in Fig. \ref{fig:coupling_evo}. The distance modulus is defined as: 
	\begin{equation}
		\mu  = 5 \log_{10}(D_L) + 25, 
	\end{equation}
	where \(D_L\) is the luminosity distance which is expressed as: 
	\begin{equation}
		D_{L} = c(1+z) \int_{0}^{z} \frac{1}{H} dz\, .
	\end{equation}
	Here, we have plotted the Hubble parameter and distance modulus \(\mu\) \footnote{Note that here \(\mu\) refers to the distance modulus, not the GB coupling parameter.} with respect to the redshift parameter \(z\). We utilized 43 OHD (cosmic chronometer) datasets \cite{Cao:2021uda} and 1701 Pantheon+ samples \cite{Pan-STARRS1:2017jku} to illustrate the evolution of trajectories corresponding to the chosen initial conditions. 
	
	The initial conditions have been selected such that during the late-time epoch, the model produces late-time cosmic acceleration and stabilizes at \(P_{6}\). We observe that for some initial conditions, the evolution of \(x\) in the far past epoch converges to either \(x = +1\) or \(x = -1\), and the corresponding potential parameter \(y\) goes to zero. The scalar field density during these epochs is nearly 1, but for some initial conditions and model parameters, it also vanishes. This vanishing nature shows that both \(x\) and \(y\) are near zero. The system's effective equation of state (\(\omega_{\rm eff}\)) exhibits a stiff matter type solution and also yields negative values. Additionally, we notice some abrupt changes in the behavior of \(\Omega_{\phi}\) and \(\omega_{\rm eff}\), which manifest in the Hubble evolution plots. However, no noticeable difference is seen in the distance modulus plots.
	
	In the future epoch, i.e., \(N \ge 0\) or \(z \le 0\), we observe that the trajectories corresponding to \(x\) converge to near zero, while \(y\) converges to 1. The scalar field density is near 1 and the effective equation of state is around \(-0.8\). Hence, for the aforementioned initial conditions and model parameter ranges, the system enters an accelerating phase with scalar field domination, describing the late-time cosmic accelerating behavior of the universe.
	
	
	
	
	\section{\ct =1: Analysis by constraining the tensor speed}
	\label{sec:const_dyn}
	
	In the previous section, we obtained the dynamics corresponding to the exponential form of the GB coupling factor and found that the model stabilizes and satisfies the gravitational waves (GWs) bound depending on certain choices of model parameters and initial conditions, particularly \(z_0\). In this section, our approach will be inclined toward model-independent analysis. As seen in the previous section, the tensor propagation speed $\ct$ depends on the time derivative of the GB coupling factor \(f\). To satisfy the model's GW speed bound, $\ct$ must be nearly equal to 1. Here, we will fix the value of $\ct$ to $1$;  consequently, the GB coupling factor \(f\) reduces to a particular form
	\begin{equation}
		\ct = \frac{1- 8 \ddot{f}}{1 - 8 H  \dot{f}} =1 \implies \dot{f} = a/f_{0},
		\label{const_gw_rel}
	\end{equation}
	where \(f_{0}\) is an integration constant having the equivalent dimension as $t$. 
	As \(\dot{f}\) is proportional to the scale factor, and \(f \equiv f(\phi)\) is a function of \(\phi\), we can express the following relations as:
	\begin{equation}
		f_{,\phi} =\frac{a}{f_{0}}\dfrac{1}{\dot{\phi}}, \quad f_{,\phi \phi} = \left(\frac{\dot{a}}{f_{0}} - \dfrac{a}{f_{0}} \dfrac{\ddot{\phi}}{\dot{\phi}}\right)\dfrac{1}{\dot{\phi}^{2}}.
	\end{equation}
	As \(f\) becomes proportional to \(a\), we must introduce a new variable to account for this change. This new dynamical variable is defined as:
	\begin{equation}
		\xi =  \frac{a H}{a_{0}H_{0} + a H },
	\end{equation}
	where \(a_{0}\) and \(H_{0}\) are constants corresponding to the present values of the scale factor and the Hubble parameter, respectively. The range of \(\xi\) becomes,
	\begin{equation}
		\xi=
		\begin{cases}
			0, \quad aH \to 0,\\
			\frac{1}{2}, \quad aH = a_0 H_0,\\
			1, \quad aH >> a_0 H_0.
		\end{cases}
	\end{equation}
	We also introduce a new constant, denoted as \(\mo\), corresponding to the Gauss-Bonnet (GB) coupling, defined by \(\mo \equiv \frac{a_0 H_0}{f_0}\). Next, we calculate \(\dot{a}\) as follows: 
	\begin{equation}
		\dot{a} = \frac{-\xi \mo f_0}{(1 - \xi)} \left(\dfrac{\dot{H}}{H^2}\right) +  \frac{\mo f_0 \xi'}{(1 - \xi)^2},
	\end{equation}
	where \(\xi' =  \frac{\dot{\xi}}{H}\). Subsequently, \(\xi'\) is determined as:
	\begin{equation}
		\xi' = \xi (1-\xi) \left( 1+ \dfrac{\dot{H}}{H^2}  \right). \label{xi_prime}
	\end{equation}
	To find \(\dot{H}\), we differentiate Eq.~\eqref{frd_eqn} with respect to time and utilize Eqs.~\eqref{dm_eqs},  \eqref{field_eos}, and 
	\eqref{const_gw_rel}, we obtain
	\begin{equation}
		\frac{\dot{H}}{H^2} \left(1 - \frac{4 \mo \xi}{1 - \xi}\right) = \frac{-3}{2} \om - \frac{4 \mo \xi }{1 - \xi} - 3 x^2 + \frac{4 \mo \xi'}{(1 - \xi)^2}.
	\end{equation}
	Substituting Eq.~\eqref{xi_prime}, we can rewrite it as follows:
	\begin{equation}
		\frac{\dot{H}}{H^2} = \frac{1}{1 - \frac{8 \mo \xi}{1- \xi}} \left(\frac{-3}{2}\om - 3 x^2\right).
	\end{equation} 
	Then, the autonomous equations for the constrained system are as follows:
	\begin{eqnarray}
		x' & =& \frac{-3 Q \om}{\sqrt{6}} + \frac{3 \lambda y^2}{\sqrt{6}} - 3x - \frac{4 \mo \xi}{(1- \xi)x} - \frac{\dot{H}}{H^2} \left(x + \frac{4 \mo \xi }{(1-\xi)x}\right), \label{x_prime_const}\\
		y' & = & \frac{-1}{2} \sqrt{6} \lambda  x y - y \frac{\dot{H}}{H^2}, \label{y_prime_const}\\
		\xi' & = & \xi (1-\xi) \left( 1+ \dfrac{\dot{H}}{H^2}  \right). \label{xi_prime_const}
	\end{eqnarray}
	
	In these autonomous equations, \(x'\) diverges as \(\xi \to 1\). However, \(\xi = 1\) represents one of the coordinates of the critical points that describe the extreme past or future epochs of the universe since \(aH \gg 1\). To study the dynamics corresponding to \(\xi = 1\), we redefine the time variable as \(dN = (1 - \xi) d\ti{N}\) to regularize the autonomous system without affecting the dynamics at other epochs. However, the Gauss-Bonnet coupling factor \(\Omega_{\rm GB} = \frac{8\mo\xi}{1-\xi}\) diverges at this point. Nevertheless, as \(\xi \to 1\) represents the extreme epochs of the universe, one can still study the system near this point. Therefore, we shall express the above autonomous equations with respect to the new time variable \(\ti{N}\) and investigate the nature of the critical points. The Friedmann constraint relation becomes:
	\begin{equation}
		\om  =  1 - x^2 -y^2 - \frac{8 \mo \xi}{(1-\xi)}.
	\end{equation}
	The critical points corresponding to the autonomous equations for the redefined time variable are summarized in Tab.  \ref{tab:critical_const_model}. The system yields seven critical points, designated with nomenclature similar to the previous case. The qualitative behavior of these critical points has been thoroughly discussed.
	
	\begin{table}[t]
		
		\centering
		\begin{tabular}{p{1.1cm}p{4.5cm}p{1.8cm}p{1.1cm}p{3.0cm} p{2.2cm}}
			\hline
			Points & \((x_*, y_*, \xi_*)\) & \(\Omega_{\phi}\)& $\Omega_{\rm GB}$ & $\omega_{\rm eff}$ & Eigenvalues\\
			\hline
			\hline
			$P_{1,2}$& $(\mp 1, 0, 0)$ & $1$ & $0$ & $1$ & Unstable \\
			\hline
			$P_{3}$ & $\left(-\sqrt{\frac{2}{3}}Q, 0,0\right)$ & $\frac{2 Q^2}{3}$ & 0 & $\frac{2 Q^2}{3}$&  Saddle\\
			\hline
			$P_{4}$ & $\left(\frac{\lambda }{\sqrt{6}}, \frac{\sqrt{-\left(\left(\lambda ^2-6\right) (\lambda +Q)\right)}}{\sqrt{6} \sqrt{\lambda +Q}}, 0\right)$ & 1 & 0 & $\frac{1}{3} \left(\lambda ^2-3\right)$ & Fig. \ref{fig:stab_p4_const} \\
			\hline
			$P_{5}$ & $\left(\frac{\sqrt{\frac{3}{2}}}{\lambda +Q}, \frac{\sqrt{Q^2+\lambda  Q+\frac{3}{2}}}{\lambda +Q},0\right)$ & $\frac{Q^2+\lambda  Q+3}{(\lambda +Q)^2}$ & 0 & $-\frac{Q}{\lambda +Q}$ & Fig. \ref{fig:exist_p5_const}\\
			\hline
			$P_{6}$ & $\left(\frac{\sqrt{\frac{2}{3}}}{\lambda }, \frac{2}{\sqrt{3} \lambda }, \frac{\lambda ^2-2}{\lambda ^2 (8 \mo+1)-2}\right)$ & $\frac{2}{\lambda ^2}$ & $1-\frac{2}{\lambda ^2}$ & $-\frac{1}{3}$ & $E_{1}<0 \rightarrow \left(\lambda>0, Q<\frac{\lambda }{2}\right)$, Fig. \ref{fig:stab_p6_const}\\
			\hline
			$P_{7}$& \(\left(-\frac{1}{2 \sqrt{6} Q}, \text{Any}, 1\right)\) & $\frac{1}{24 Q^2}+y_*^2$ & $-\frac{8 M \xi _*}{\xi _*-1}$ & $-\frac{\left(\xi _*-1\right) \left(24 Q^2 y_*^2-1\right)}{24 Q^2 \left(8 \mo \xi _*+\xi _*-1\right)}$ & Fig. \ref{fig:stab_point7_const1}\\
			\hline
		\end{tabular}
		\caption{The critical points corresponding to the constrained system $\ct =1$.}
		\label{tab:critical_const_model}
	\end{table}
	
\begin{figure}[t]
		\centering
		\includegraphics{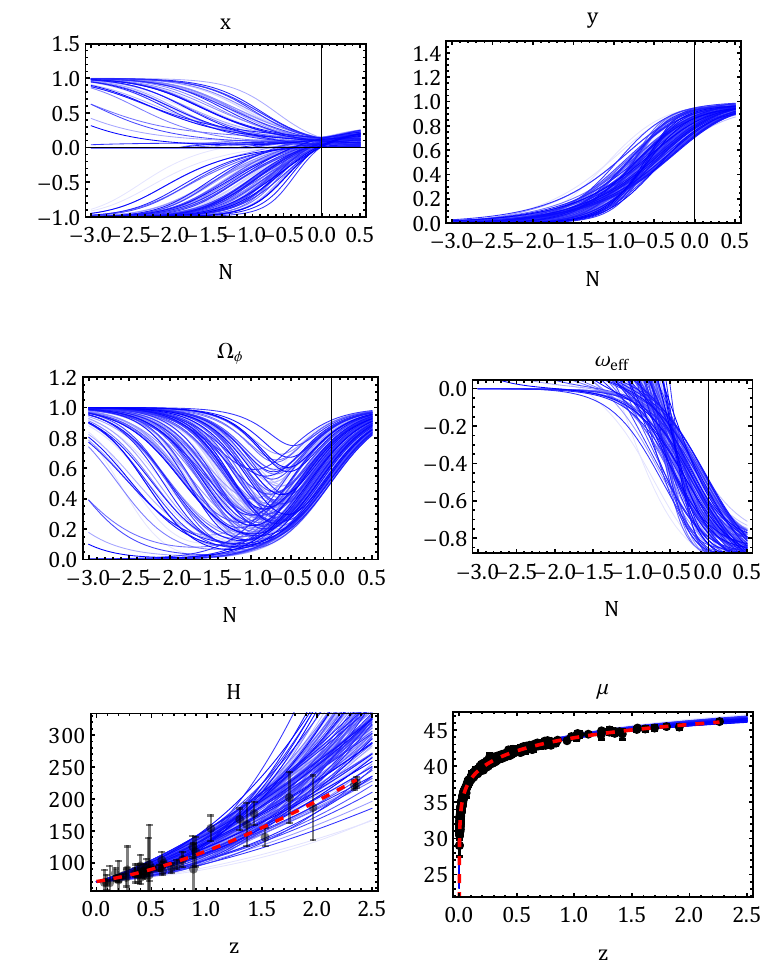}
		\caption{The evolution of dynamical variables corresponding to the interacting case for \(H_0 = 70 \ \rm Km \ s^{-1} Mpc^{-1}\). Here, the dotted red line is plotted for a flat $\Lambda$CDM model. For $H$ vs $z$ plot, the best value corresponding to flat $\Lambda$CDM model has been used \(H_0 = 70.6  \ \rm Km\ s^{-1} Mpc^{-1},\) \(\Omega_\Lambda = 0.74\), however, for distance modulus($\mu$), \(H_0 = 73.6  \ \rm Km \ s^{-1} Mpc^{-1}\), and \(\Omega_\Lambda=0.67\). }
		\label{fig:entire_evo_constrained}
	\end{figure}

	\begin{itemize}
		
		\item \textbf{Points $P_{1,2}$:} These points are dominated by the kinetic part of the field, \(x_* \ne 0\), while the other coordinates vanish. The field density parameter dominates over the fluid and GB coupling density, resulting in ultra non-accelerating dynamics with $\omega_{\rm eff} = 1$. The points always exhibit unstable behavior.

		\item \textbf{Point $P_{3}$:} Similar to the unconstrained scenario discussed previously, the point's kinetic coordinate is non-zero and depends on the coupling parameter \(Q\). This point exists for \(Q \ne 0\) and signifies the non-accelerating past regime of the universe. The point also exhibits unstable behavior.
		
		\item \textbf{Point $P_{4}$:} This point is dominated by the field energy density and produces an accelerating solution for \(0 \le |\lambda | \le 1.22\). The stability of the point has been depicted in Fig. \ref{fig:stab_p4_const}. However, stability is achieved only for $\lambda\ge 1.40$. Consequently, this point cannot be considered a viable fixed point.
		
		\begin{figure}[t]
			\centering
			\includegraphics[scale=0.6]{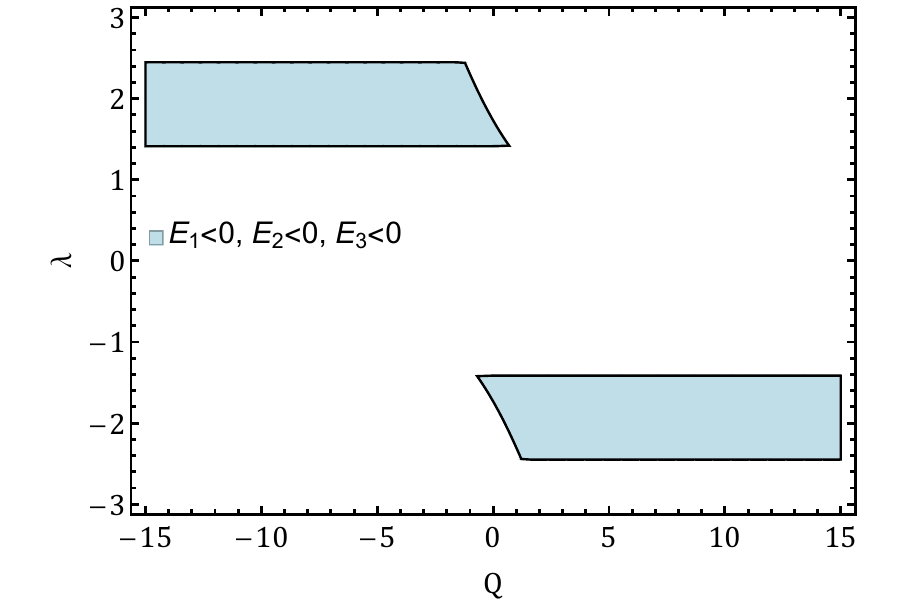}
			\caption{The stability of the point $P_{4}$.}
			\label{fig:stab_p4_const}
		\end{figure}
		
		\item \textbf{Point $P_{5}$:} The existence and stability of this point have been depicted in Fig. \ref{fig:exist_p5_const}. Considering all the necessary constraints, it is evident that the point cannot yield a stable accelerating solution in the late-time epoch.
		
		\begin{figure}[t]
			\centering
			\includegraphics[scale=0.6]{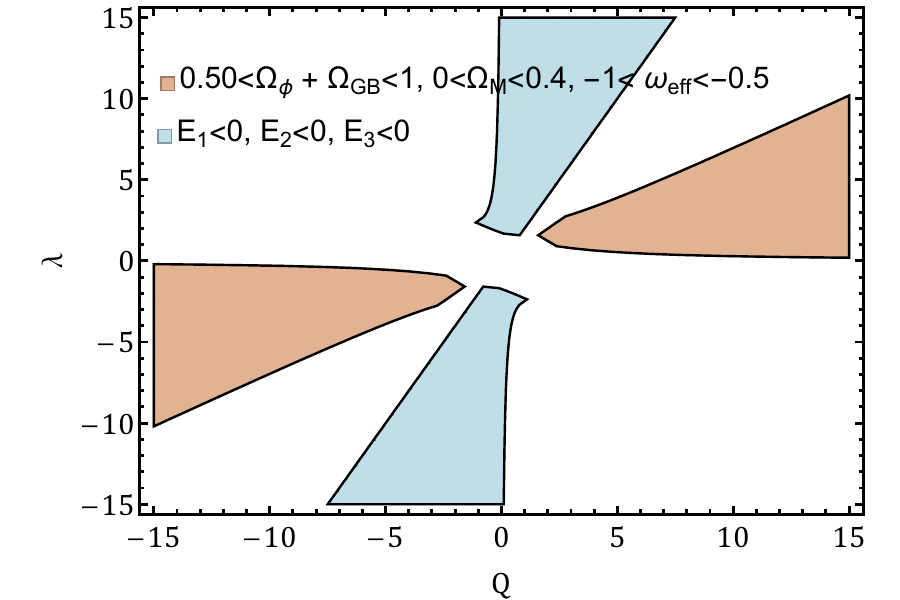}
			\caption{The field and fluid energy density domination and acceleration effective EoS range with stability criteria corresponding to the point $P_{5}$.}
			\label{fig:exist_p5_const}
		\end{figure}
		
		\item \textbf{Point $P_{6}$:} At critical point $P_{6}$, $\xi$ and $\Omega_{\rm GB}$ are non-zero, yet the effective equation of state (EoS) is $-1/3$, markedly deviating from the concordance model of cosmology. Stability analysis reveals that the real parts of the eigenvalues are negative for \(0 \le |\lambda| <\sqrt{2}\) as shown in Fig. \ref{fig:stab_p6_const}. In this scenario, the field density $\Omega_{\phi}>1$. However, investigating the system's evolution corresponding to the benchmark points where stabilization occurs at this point fails to yield a consistent matter-dominated epoch, nor does the effective EoS approach \(-1\) during the current epoch. Consequently, $P_{6}$ cannot be considered a physically viable point to describe various phases of the universe.
		
		\begin{figure}[t]
			\centering
			\includegraphics[scale=0.6]{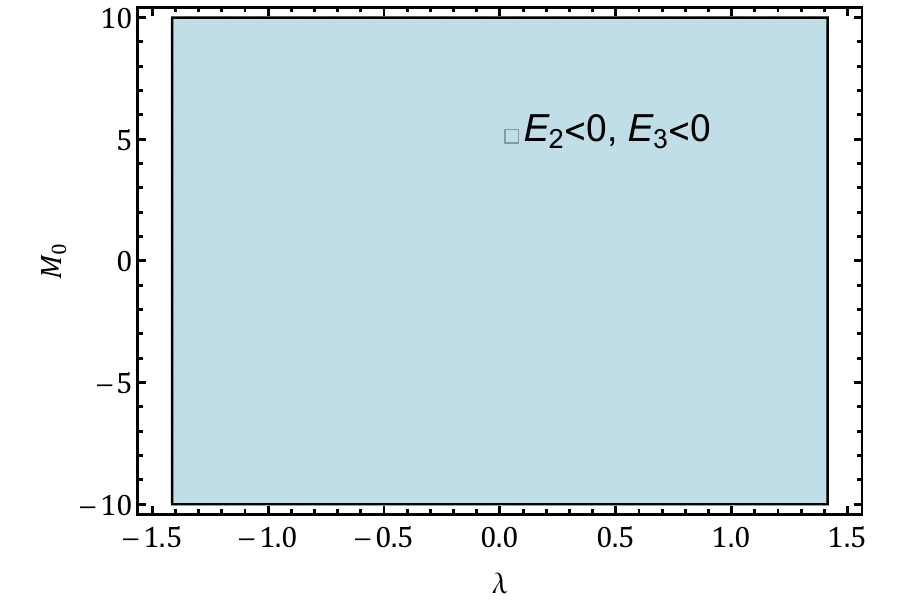}
			\caption{The nature of eigenvalues corresponding to the point $P_{6}$.}
			\label{fig:stab_p6_const}
		\end{figure}
		
		\item \textbf{Point $P_{7}$:} Point $P_{7}$ exists for \(Q \ne 0\) and can take any value of \(y\). At this point, the field density depends on the interaction parameter \(Q\) and the value of \(y\). Thus, we select \(Q\) and \(y\) such that the fractional field density satisfies \(0 < \Omega_{\phi} \leq 1\). Additionally, we consider that the system produces an accelerating solution when \(y > x\), thus one can choose \(y_* = (0.8-- 0.9)\). The effective equation of state becomes \(\mo\), \(Q\), \(y\), and \(\xi\) dependent. For $\xi \to 1$, we observe that $\omega_{\rm eff} \to 0$. However, since we are not interested in the exact value of \(\xi = 1\), we analyze the system's behavior near \(\xi = 1\) and present the analysis for \(\xi_* = 0.99\). Assessing the stability of the point in the parameter space of \((Q, \mo)\) by fixing \(y_*, \xi_*\), as shown in Fig. \ref{fig:stab_point7_const1}, we find that the system imposes a tight constraint on $\mo$: \(-10^{-5} < \mo < 0\). The GB constant factor $\mo$ plays a significant role in keeping $\Omega_{\rm GB} \to 0$. For larger negative values of $\mo$, $\Omega_{\rm GB}$ diverges rapidly, rendering the system unstable in the near future epoch.
		
		\begin{figure}[t]
			\centering
			\includegraphics[scale=0.6]{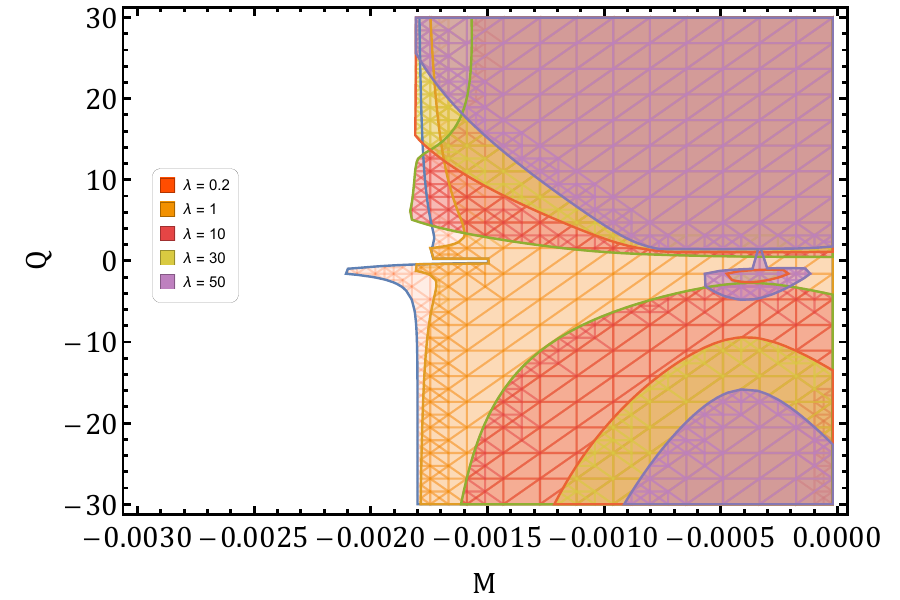}
			\caption{The stability corresponding to the point $P_{7}$ for \(y_{*} = 0.9, \xi_{*} = 0.99\).}
			\label{fig:stab_point7_const1}
		\end{figure}
		
	\end{itemize}
	From the above examination of the behavior of the fixed points, it becomes clear that out of the seven critical points, only $P_{7}$ can be a physically viable point that describes the late-time state of the universe. To explicitly demonstrate the dynamics of the system, we illustrate the evolution of the cosmological parameters corresponding to the interacting case \(Q \ne 0\) in Fig. \ref{fig:entire_evo_constrained}, by generating random numbers within the range of initial conditions \(0.01\le x_0 \le 0.1, \; 0.7 \le y_0 \le 0.95, \xi_0 = 0.5, \; 0.01 \le \lambda \le 0.8,\; \mo = -10^{-7}, \; -0.2 \le Q \le -0.001 \). We have fixed certain parameters, such as $\mo$ and $\xi_0$ while varying the rest of the parameters to ensure that the system generates an accelerating solution and the scalar field density parameter dominates during the current phase. Note that we have evaluated all the parameters with respect to the original time variable \(N\), not \(\ti{N}\), since we are not exploring the far state of the universe. As stated earlier, the redefined autonomous system only regularizes the extreme past or future state of the system, not the intermediate dynamics. Hence, evaluating the system's dynamics corresponding to its original time variable \(N\) won't produce different dynamics than \(\ti{N}\). We have also plotted the Hubble parameter and distance modulus, \(\mu\), against redshift, \(z\), and observed that corresponding to the Hubble evolution, at low redshifts, \(z< 0.8\), most of the trajectories fall within the range of the data points; however, as the redshift increases, the curves start to deviate from the data points. Here, we observe in the evolution of dynamical variables that in the past epoch, similar to the previous case, \(x\) converges to \(\pm 1\), \(y\) goes to zero, and the field density dominates. The effective equation of state is either near zero or deviates from zero, depending on the values of \(x\). At the current epoch, the model transitions to the accelerating phase with the domination of scalar field density.
	
\begin{figure}[]
		\centering
		\includegraphics[scale=0.5]{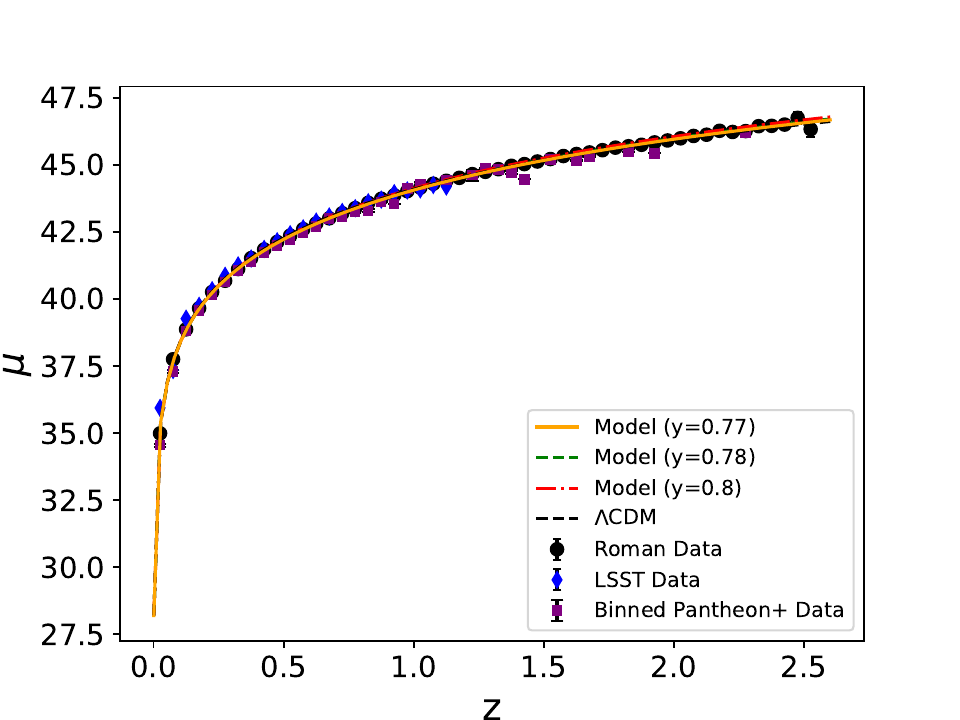}
		\hspace*{2cm}	\includegraphics[scale=0.5]{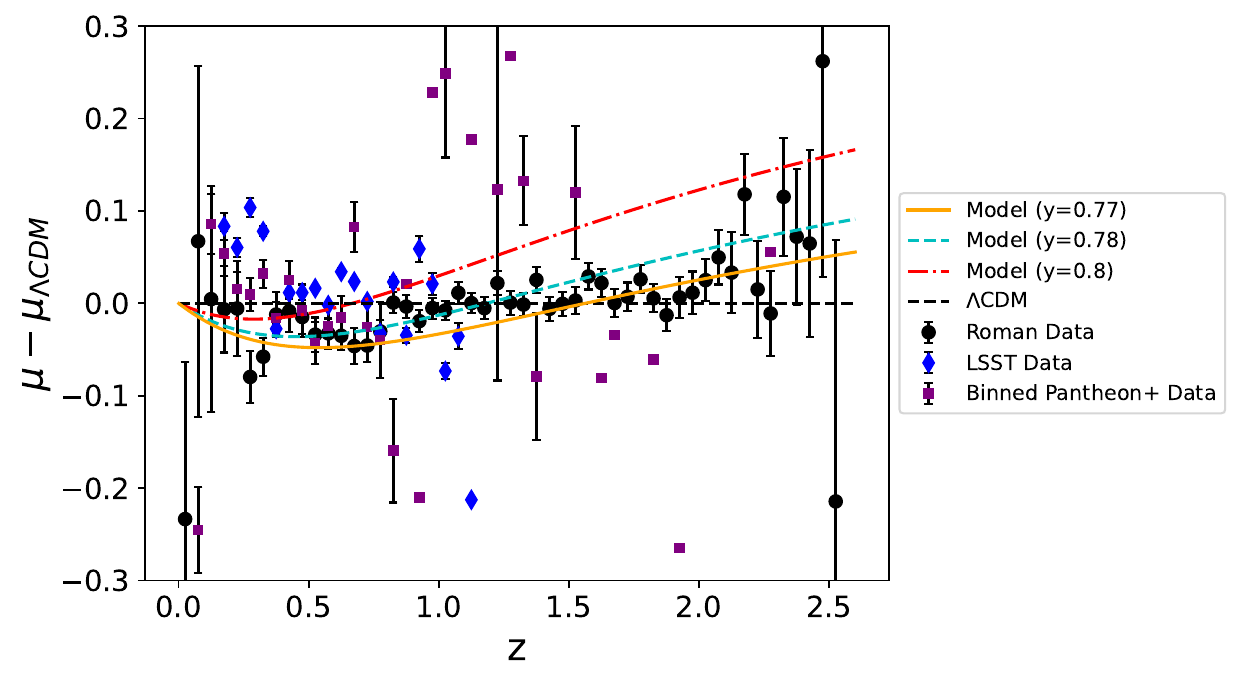}         
		\caption{ The top panel shows the distance modulus $\mu$ versus redshift $z$. The different surveys are each given different colors. The bottom panel is the distance-modulus residuals relative to a best-fit cosmological model with each binned data. Both the data errors and the binned data errors include only statistical uncertainties. At $z <0.01$, the sensitivity of peculiar velocities is very large and reflects this uncertainty. We use the initial conditions as $x_{0}=0.01$, $\xi=0.5$, $M_0 = -10^{-7}$, $Q = -0.001$, $\lambda = 0.01$. {It can be seen that the uncertainties and redshift range of the LSST and Roman simulated data are such that a significantly larger range of parameter space can be observed verses $\Lambda$CDM. This is especially true at $z>0.75$.}}
		\label{m}
	\end{figure}

	In the next section, we will present the check of our model with a different set of datasets. The complete SNIa sample utilized to measure cosmological parameters from the three different telescope data is depicted in Fig. \ref{m}. It displays the binned distance residuals from SNIa and the binned distance residuals from all three SN samples. At a given redshift, foundation SNIa exhibits greater average distances and a corresponding positive shift in Hubble residuals compared to SNIa from the earlier low-z sample. The highest and lowest redshift points have extremely high uncertainties, as no SNe are above or below them in redshift. It is worth noticing that the Pantheon+ samples are more noisy than the other two datasets. At the low redshift end, the redshift-dependent information from Roman galaxies and its cross-correlation allows constraints on deviations in the growth history in a redshift-dependent way. We see that the Roman data leads to significantly improved constraints, especially at high redshift. 
	
	\section{Datasets and methodology}\label{sec:MCMC}
	In the present section, we summarize the datasets used in our analysis below.
	\begin{itemize}
		\item \textit{Hubble parameter:} Values of the Hubble parameter $H(z)$ are usually derived from the so-called differential age of the galaxies (DAH) methodology. As the Hubble parameter can be estimated at a redshift $z$ from $H(z)=\frac{-1}{1+z}\frac{dz}{dt}$, $dz/dt$ can be obtained from measurements of massive and very slowly evolving galaxies, dubbed Cosmic Chronometers (CC). We use 31 points compiled in \cite{Yu_2018}, where we take them as uncorrelated.
		
		\item \textit{The Pantheon+ dataset:} It comprises distance moduli calculated from 1701 light curves of 1550 SNIa \cite{Brout_2022,Murakami_2023}. These light curves were obtained from 18 different surveys, and they cover a redshift range of $0.001 \leq z \leq 2.26$. It is worth noting that 77 out of the 1701 light curves are associated with Cepheid-containing galaxies. We refer to this combined dataset as Pantheon+SH0ES.
		
		\item \textit{{Simulated} LSST survey:} The Vera C.~Rubin Observatory Legacy Survey of Space and Time (LSST) is anticipated to process around $10^6$ transient detections per night \cite{LSSTDarkEnergyScience:2021laz}. For accurate measurements of cosmological parameters and rates, it is essential to understand the detection efficiency, magnitude limits, levels of artefact contamination, and biases in the selection and photometry. The LSST is projected to increase the sample size of Type Ia supernovae by up to 100 times compared to previous studies \cite{FSS:2018cey,DES:2017dgt}. Moreover, the survey will enable the discovery of SNeIa using a single instrument, with redshifts reaching up to about $1.2$. The systematic uncertainty requirements for the SNIa cosmology analysis, as outlined by The LSST Dark Energy Science Collaboration et al. \cite{LSSTDarkEnergyScience:2018jkl}, include achieving photometric precision at the few-millimagnitude level and accurately determining selection biases. It is reasonable to approximate the LSST data set as 25 times larger \cite{LSSTDarkEnergyScience:2018jkl,Vincenzi2023} than the DES 5Y sample \cite{Sanchez2024} with a 15 times larger low-z sample \cite{LSSTDarkEnergyScience:2018jkl}.
		
		\item \textit{{Simulated} Roman survey}: The Nancy Grace Roman Space Telescope is a NASA flagship mission with a diverse portfolio of science goals, including cosmology, exoplanet research, and general astrophysics. One of the major components of the mission is the High Latitude Wide-Area Survey, which includes both imaging and spectroscopic elements \cite{Wang:2021oec}. Roman's planned High Latitude Time-Domain Survey enable the detection and light-curve monitoring of thousands of Type Ia supernovae up to $z \approx 3$ \cite{Hounsell2018,Rose:2021nzt,Hounsell:2023xds}. The combination of high-resolution imaging and spectroscopy for precise brightness and redshift measurements will reduce the uncertainty in the dark energy equation of state by 70\%, quantify temporal variations, and identify inflection points in the recent expansion rate. 
		
		In the current reference survey for the High Latitude Time-Domain survey \citep{Rose:2021nzt}, the Roman Space Telescope will observe the same part of the sky every five days for two years. This will be done in two tiers, one that cover an area of $\sim$5~$deg^2$ to a depth of 26.5~mag per visit, and another that covers $\sim$20~$deg^2$ to a depth of 25.5~mag. In this setup, Roman is expected to observer over 12,000 cosmologically useful SNIa, as well as many more other transient types. For now, we are able to simulate this data set by using the simulation components of the \texttt{SNANA} software \citep{Kessler2009} with the \texttt{PIPPIN} \citep{Hinton2020} pipeline manager {and assuming $\Lambda$CDM}. The details for the Roman simulations came from communications about a paper in prep.~from the Roman Supernova Project Infrastructure Team. The simulation input values will be released with that forthcoming paper, but are derived from the information presented publicly from the Roman project at \url{https://roman.gsfc.nasa.gov/science/WFI_technical.html}.
		
	\end{itemize}
	
\begin{figure}[H]
		\centering
		\includegraphics[scale=0.47]{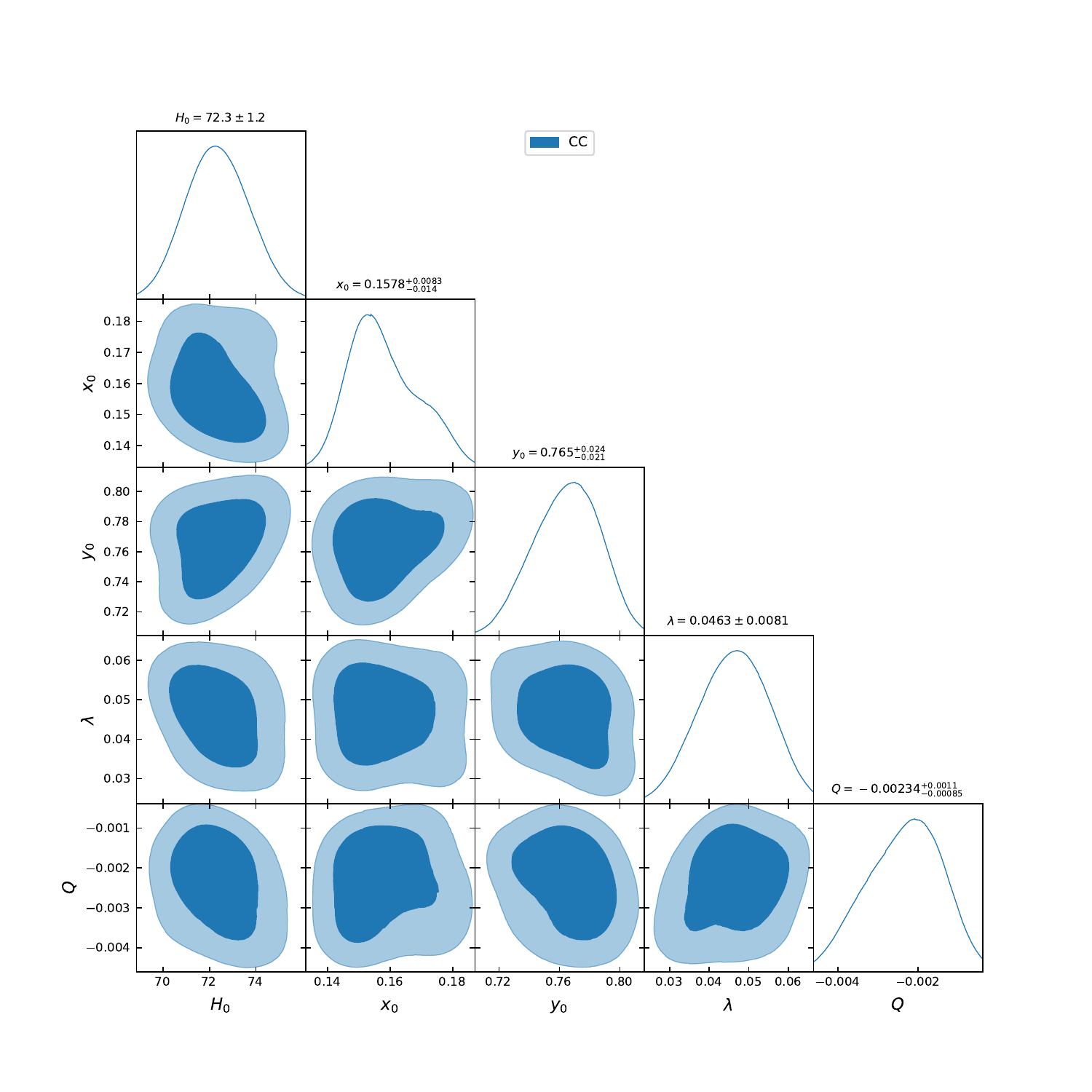}
		\caption{Marginalized posterior distributions and contours (68\% and 95\% CL) using CC data for the case $c_{T}^2 =1$.}
		\label{CC}
	\end{figure}

\begin{figure}
		\centering
		\includegraphics[scale=0.47]{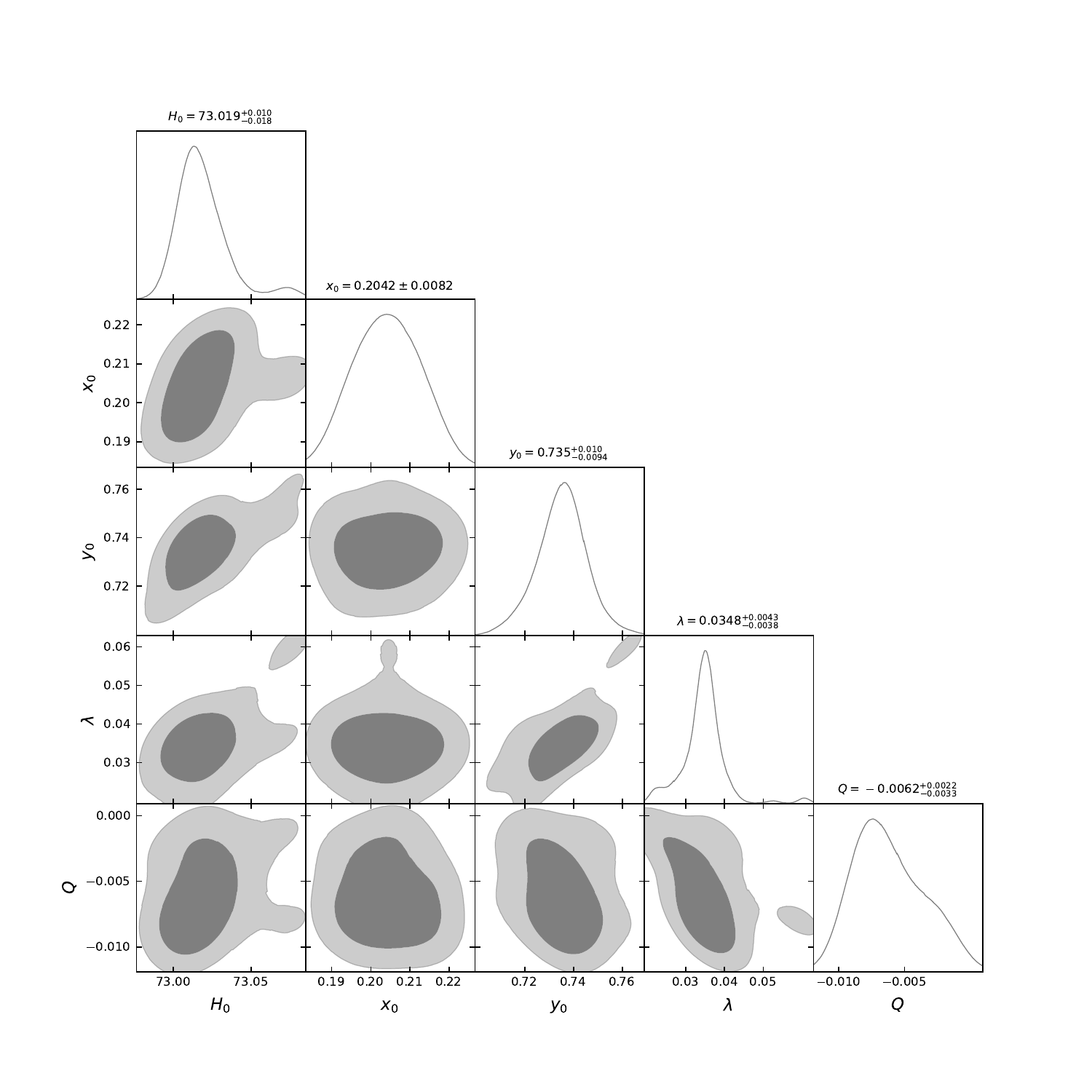}
		\caption{Marginalized posterior distributions and contours (68\% and 95\% CL) using Pantheon+ data for the case $c_{T}^2 =1$.}
		\label{SN}
	\end{figure}

	Throughout our statistical analyses, we adopt flat priors for all parameters: $H_{0} \in [60, 100]$, $x_{0} \in [0.13, 0.25]$, $y_{0} \in [0.7, 0.9]$, $\lambda \in [0.01, 0.07]$, $Q \in [-0.001, -0.01]$.  We employ Monte Carlo Markov Chain (MCMC) techniques to sample the posterior distributions of the model's parameters. We also present in Figs. \ref{CC} and \ref{SN} the one and two-dimensional marginalized distributions of the extended model parameters at 68\% and 95\% CL for the CC and Pantheon+SH0ES datasets. We observe a strong correlation between $H_0$ and other parameters. With that being said, all the datasets that favor somewhat large $H_0$ values, with ($> 70$), also show a preference for a relatively significant deviation. Using distances and a stat+syst covariance matrix that extends to the Cepheid calibrators and combining the Pantheon+ SNe with the SH0ES Cepheid host distance calibration, we are able to robustly and simultaneously constrain $H_0$ and other cosmological parameters describing the expansion history. It is noted that the values of $H_{0}$ is in excellent agreement with the SH0ES measurement of $H_0 = 73.04 \pm 1.04\,km\,s^{-1}\,Mpc^{-1}$.
	
	For the statistical comparison, the AIC difference between the model under consideration and the reference model is calculated. This difference in AIC values can be interpreted as evidence in favor of
	the model under consideration over the reference model. It is useful to estimate how preferred the proposed models are in comparison with the standard $\Lambda$CDM one. In case of $\Lambda$CDM, we obtain $H_{0}= 68.84^{+0.090}_{-0.13}$ and $\Omega_{m0} = 0.315\pm 0.029$ for CC samples whereas $H_{0}=73.03\pm 0.0093$ and $\Omega_{m0}=0.372 \pm 0.0062$ for SN samples.  We then incorporate statistical criteria, the Akaike Information Criterion (AIC) and Bayesian Information Criterion (BIC), defined as \citep{1100705}, $\mathrm{AIC} = \chi^2_{\mathrm{min}}+2d$ and $\mathrm{BIC} =\chi^2_{\mathrm{min}}+d\ln N$, where $d$ is the number of free parameters, and $N$ is the total size of the data. \\

	\begin{table}[H]
		\renewcommand{\arraystretch}{1.3}
		\setlength{\tabcolsep}{19pt}
		\begin{center}
			\caption{The corresponding $\chi^{2}$ of the models for each sample and the information criteria AIC and BIC for the examined cosmological model.}
			\label{Table}
			\begin{tabular}{|l c c c c|}
				\hline\hline
				& Data & $\chi^{2}_{min}$ & $AIC$  & $BIC$   \\ \hline \hline
				Model & CC &  $19.48$  &   $29.48$ &  $36.6$   \\ 
				& SN   & $1664.5$ & $1674.5$ & $1701.7$     \\
				\hline \hline
				$\Lambda$CDM & CC &  $16.06$  &  $20.06$ &  $22.9$   \\
				& SN  & $1667.2$ & $1673.2$ &  $1689.5$  \\
				\hline \hline
			\end{tabular}
		\end{center}   
	\end{table}
	
	In this criterion, if the difference in AIC value between a given model and the best one ($\Delta$AIC) is less than 4, both models are equally supported by the data. For $\Delta$AIC values in the interval $4 < \Delta AIC < 10$, the data still support the given model but less than the preferred one. For $\Delta AIC > 10$, the observations do not support
	the given model. Similarly, BIC discriminates between models as follows: For $\Delta BIC< 2$, there is no appreciable evidence against the model. If $2 < \Delta BIC< 6$, there is modest evidence against the considered model. For the interval $6 < \Delta BIC< 10$, the evidence against the candidate model is strong, and even stronger evidence against it exists in the data when $\Delta BIC> 10$.
	
	{Additionally}, we highlight that Pantheon+SH0ES show clear and strong evidence in favor of our considered model (Refer to Table \ref{Table}). For clarification purposes, the interpretation follows from the approach of $\Delta$AIC values. The model above performs significantly better than $\Lambda$CDM from the point of view of data. 
	
	{Finally, we introduce the usefulness of the future data from LSST and Roman. With the increased redshift coverage, out to $z \sim 2.5$, and distance precision reduced by an order of magnitude, they will reveal more about the evolving dark energy. For these classes of models, the deviation from $\Lambda$CDM is significant at $z > 0.5$ and grows largest at the highest redshifts---right where these new surveys are optimized to collect data. We look forward to using these datasets in the 2030s to further constrain models beyond $\Lambda$CDM.
	}

	

	\section{Conclusion}
	\label{sec:conclusion}
	
	In this paper, we have discussed the Gauss-Bonnet (GB) coupled dark energy model, specifically the quintessence scalar field interacting with the dark matter fluid via the interaction term ${\cal{Q}}$, introduced at the covariant derivative of the energy-momentum tensor. One of the notable features of GB-coupled models is the varying speed of gravitational waves (GWs). However, recent detections of GWs and $\gamma$-rays have imposed a stringent constraint on the GW speed, specifically $|\ct-1|<  5\times 10^{-16}$. 
	
	This study focuses on analyzing the dynamics of the model using the dynamical system technique under two primary scenarios. Initially, an exponential GB coupling was introduced along with the interacting scalar field and fluid. In this scenario, $c_T$ is allowed to vary and the corresponding dynamical system analysis was set up, resulting in a 3-dimensional phase space. To constrain the dimension of the phase space to less than 3 dimensions, we chose the interaction term \(\mathcal{Q} \propto Q \rho_{m} \dot{\phi}\). Generally, the choice of interaction term tends to increase the dimensionality of the phase space, making the analysis mathematically challenging. Critical points corresponding to this dynamical setup were obtained, signifying different phases of the universe. For instance, some critical points represent the matter phase, while others yield an effective equation of state (EoS) of $-1$, signifying the late-time de-Sitter phase. Some critical points indicate a late-time accelerating solution with varying effective EoS, primarily depending on model parameters. Evaluating $\ct$ at these points reveals no deviation between the GW speed and the speed of light. However, numerical evolution requires fine-tuning, particularly the value of \(z(0) \sim 10^{-17}\), corresponding to the GB coupling parameter, to ensure the model remains constrained to the above GW bound in the past epoch. Thus, the gravitational wave constraint implies that the GB coupling parameter \(f_{,\phi} \sim 10^{-19}\). 
	
	In the second scenario, we began the analysis by constraining $\ct=1$, which immediately restricts the form of \(f\). The time derivative of $f$ becomes proportional to the scale factor \(a\), i.e., \(\dot{f} \propto a\). We analyzed this particular scenario using the dynamical system technique. It is noteworthy that although the form of the GB coupling is constrained, closing the dynamical variables that qualify the autonomous system of equations still requires three independent variables, resulting in a 3-dimensional phase space. The system yields critical points representing different phases of the universe, with one critical point becoming asymptotically stable, qualifying to describe the late-time state of the universe. 
	
	We have conducted a study in light of observational data, focusing on implementing our considered model for the case of $c_{T}^{2}=1$. This analysis utilized apparent magnitude-redshift data from the Pantheon+ sample, which includes 1701 light curves of 1550 distinct SNIa spanning a redshift range of 0.001 to 2.26. In particular, we have managed to constrain the free model parameters using CC and SNIa. It is noted that the values of $H_{0}$ is in excellent agreement with the SH0ES measurement of $H_0 = 73.04 \pm 1.04\,km\,s^{-1}\,Mpc^{-1}$. The comparison with the observational data on the distance modulus indicates a very good
	concordance between the model and $\Lambda$CDM and observations
	up to a redshift of $z \approx 1$, with some deviations appearing
	at higher redshifts. 
	
	The AIC analysis also confirms the existence of mild tension between the present model and the $\Lambda$CDM predictions, but to obtain a definite answer to this question, more observational data spreading on a larger redshift range is necessary. 
	Incorporating higher-order terms allows for creating models that accurately fit supernova (SN) data and develop potentials similar to those in $\Lambda$CDM. Moreover, the Roman telescope has the capability to restrict other models beyond $\Lambda$CDM. {Via simulations,} we have examined Roman's abilities and utilized binned LSST data to check the constraining power of a variety of interesting extensions. Our future plans include further investigation of cosmological fitting with these datasets once the data is obtained and released over the next decade.
	
	{Finally, we would like to note that the important issue regarding to the ghost-free and Laplacian stability in the presence of the interaction between DE and DM is not addressed in this paper. Currently, we are working on the interacting models from the point of view of particle physics, in which the DM is described by a scalar/femionic field. In such models, this issue can be properly addressed, and we wish to report such studies soon.}
	
	\begin{acknowledgments}
		Y.R. is supported through Baylor Physics graduate program. B.R. is partially supported as a member of the Roman Supernova Project Infrastructure Team under NASA contract 80NSSC24M0023.  A.W. is partly supported by the US NSF grant, PHY-2308845. This work was completed, in part, with resources provided by the University of Chicago's Research Computing Center.
	\end{acknowledgments}

	\bibliographystyle{JHEP}
	\bibliography{ref}

\end{document}